# Transient Depth Thermography for Probing Heat Transport

Dmitrii Shymkiv[1], Chun Wang[1], Yan Jiang[1], Rajat Srivastava[2], Nandika D'Souza[2], and Yuzhe Xiao[1*]

[1]*Department of Physics, University of North Texas, Denton, TX 76203, USA*
[1]*Department of Mechanical Engineering, University of Texas at Dallas, Richardson, TX 75080, USA*
**Email address: yuzhe.xiao@unt.edu*

**Abstract**

Directly probing heat propagation inside materials remains challenging because conventional measurements are predominantly sensitive to surface temperature. Depth thermography has enabled non-contact reconstruction of subsurface temperature profiles from spectrally resolved thermal radiation under steady-state conditions. Here, we extend this approach into the time domain, establishing transient depth thermography to resolve the evolution of internal temperature during heat transport. By exploiting wavelength-dependent optical penetration depth, time-resolved thermal-radiation spectra provide access to temperature as a function of both depth and time. Tracking this spatiotemporal temperature field enables direct probing of heat propagation and quantitative determination of out-of-plane thermal conductivity and interfacial thermal resistance. We demonstrate the approach in fused silica, obtaining thermal conductivity within 2% of established values, and measure the temperature-dependent thermal conductivity of $MgF_2$ over a broad temperature range where existing data are sparse and inconsistent. Numerical simulations further demonstrate its extension to multilayer thin films for probing interfacial thermal resistance. By extending depth-resolved thermal spectroscopy from steady-state to transient heat transport, this work establishes a new optical route for non-contact characterization of thermal dynamics in bulk and layered materials.

## INTRODUCTION

Light provides a powerful, non-contact means of probing matter and its physical properties. Among these, thermal properties and temperature are of particular importance, yet conventional optical thermography remains largely limited to measurements at or near the surface [1,2]. Accessing the thermal state beneath a surface is particularly challenging because the internal temperature distribution is generally inaccessible to direct measurement. Depth thermography recently overcame this limitation by exploiting wavelength-dependent optical penetration in semitransparent materials, enabling the reconstruction of subsurface temperature profiles from spectrally resolved thermal radiation under steady-state conditions [3,4]. Depth thermography provides optical access to temperature beyond the surface, but its application to heat transport remains fundamentally limited by the steady-state nature: thermal transport is inherently a dynamical process, and the temporal evolution of the internal temperature field contains direct information about how heat propagates through a material.

This limitation is particularly important for thermal-transport measurements in thin films and multilayer structures, where out-of-plane thermal conductivity is critical but difficult to determine accurately [5–7]. Existing techniques, such as 3ω [8–10], thermoreflectance [6,11–14], and laser flash [15–18] generally infer thermal transport from electrical or optical signals measured at the surface and rely on models of heat propagation and boundary conditions. Such indirect approaches can become increasingly challenging for heterogeneous structures with buried interfaces, where interfacial thermal resistance, anisotropic heat transport, and uncertainties in layer thicknesses and interface geometry complicate the interpretation [1,19–21]. Direct access to the internal temperature field could therefore enable a fundamentally different approach to thermal-transport characterization, with significant advantages over conventional methods.

Here, we extend depth thermography from steady-state temperature mapping to transient heat transport. We introduce transient depth thermography, in which time-resolved thermal-radiation spectra are used to reconstruct the temperature field as a function of both depth and time. Because different wavelengths probe different depths according to their optical penetration depths, the temporal evolution of the thermal-radiation spectrum encodes the propagation of heat through the material. This depth- and time-resolved temperature field provides direct access to transient heat transport and enables quantitative determination of thermal properties, such as out-of-plane thermal conductivity and interface thermal resistance. Unlike conventional thermal-transport measurements that infer heat propagation from surface observables, transient depth thermography directly tracks the evolution of the internal temperature distribution.

We validate the technique using a fused-silica window, obtaining excellent agreement with established thermal conductivity, and further demonstrate its capability by measuring the temperature-dependent thermal conductivity of $MgF_2$. Numerical simulations further demonstrate its applicability to multilayer thin films, enabling the characterization of interfacial thermal resistance. With a measurement uncertainty of 2–7%, transient depth thermography provides a noninvasive and broadly applicable approach for thermal characterization of bulk and thin-film materials, particularly systems in which buried interfaces or complex thermal pathways limit conventional measurements.

## Results

The working principle of transient depth thermography is illustrated in Fig. 1. When a room temperature sample is placed on a preheated stage, heat is introduced from the bottom surface, generating a transient, depth-dependent temperature gradient that evolves over time. Such a transient temperature gradient results in a corresponding transient thermal-radiation spectra emitting from the top surface, which is measured using an infrared (IR) spectrometer.

The key idea is that thermal-radiation spectrum within the semitransparent spectral region carries information not only about surface temperature, but also internal temperature at different depth within the material [3]. Different depth-dependent temperature profiles produce distinct spectral signatures, which are determined by the optical properties of the material (i.e., the penetration depth). By analyzing the temporal evolution of these spectra, one can reconstruct the temperature distribution as a function of both depth and time (Fig. 1d,e). The transient depth-resolved temperature profile directly encodes the heat transport process and, therefore, enables determination of thermal properties, such as thermal conductivity and interface thermal resistance.

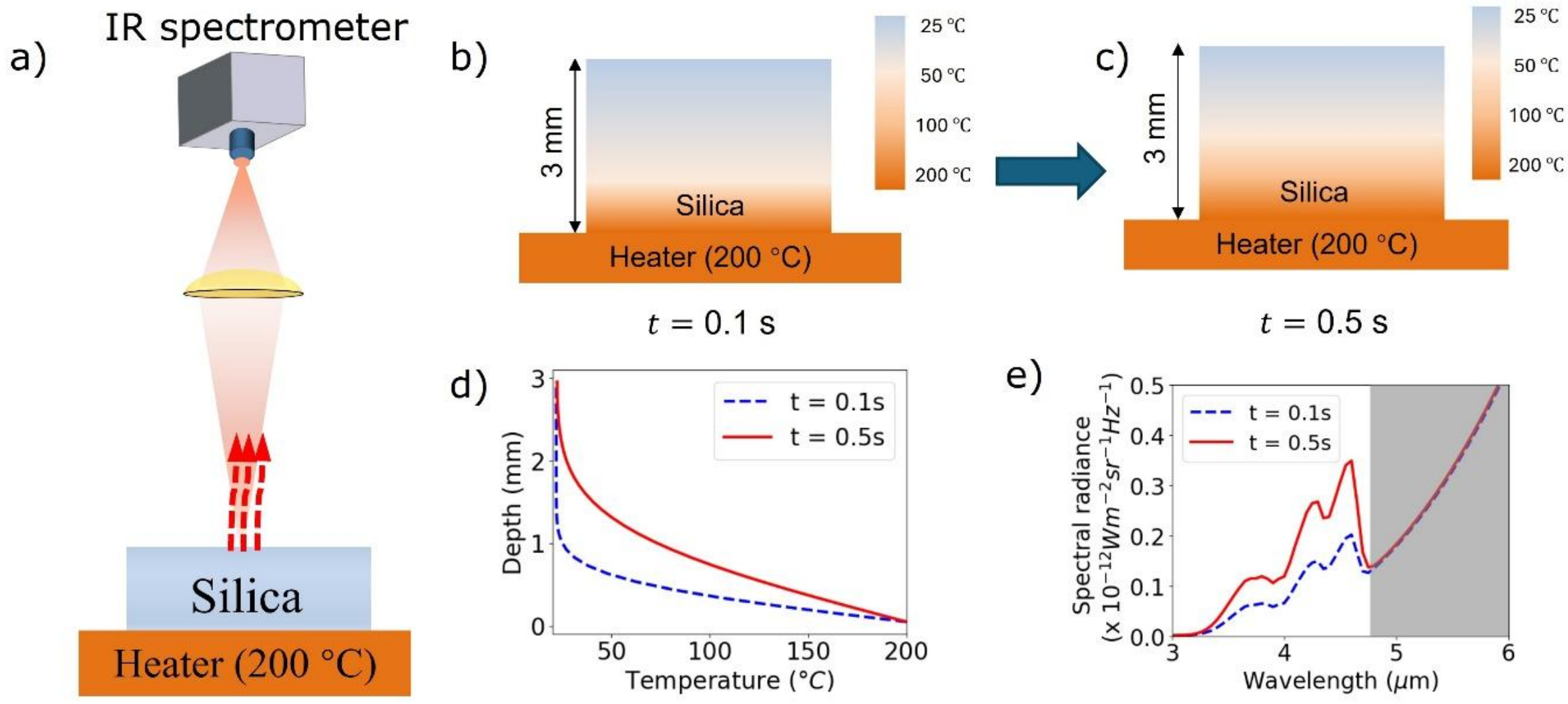

**Fig.1.** Concept of transient depth thermography. (a) A room temperature sample (fused silica as an example here) is placed onto a heater, and its transient thermal-radiation spectra are measured by an IR spectrometer. (b-c) Simulated temperature profile for a 3 mm thick fused-silica window at t = 0.1 and 0.5 seconds after placing on a 200 °C heater surface. (d) Depth-dependent temperature gradient at t = 0.1 s and 0.5 s. (e) The corresponding thermal-radiation spectra. Spectrum in the opaque (shown as shaded area) spectral range only depends on surface temperature, while spectrum in the semitransparent spectral range reflects internal temperature distribution.

To illustrate the concept, we consider a 3-mm-thick fused silica window. Fused silica exhibits strongly wavelength-dependent absorption: it is highly transparent from the visible to the near-infrared (up to ~3.5 μm) but becomes opaque in the mid infrared (above ~5 μm), leaving a semitransparent spectral window between them. This spectral dependence enables thermal radiation at different wavelengths to probe different depths within the material. For a 3-mm-thick sample, the semitransparent spectral region lies near 4 μm, where the optical penetration depth is on the order of millimeters, comparable to the sample thickness (Supplementary Note 1). Consequently, thermal radiation within this spectral window contains depth-dependent temperature information from throughout the sample, rather than only from the top surface.

Figures 1d and 1e show simulated temperature profiles and the corresponding thermal-radiation spectra emitting from the top surface at two different times in the heating process. In the opaque spectral range ($\lambda > 5$ μm), emitted thermal radiation only comes from the surface, resulting in identical spectra (because of equal surface temperature) despite quite different internal temperatures. In contrast, in the semitransparent range ($\lambda < 5$ μm), radiation emerges from a finite depth within the material. Consequently, the emitted spectra vary strongly with the internal temperature distribution. This spectrum-induced depth selectivity enables direct retrieval of depth-resolved temperature profiles.

When applied to transient heat transport, depth thermography provides direct access to the spatiotemporal evolution of temperature inside a material. Rather than inferring thermal properties solely from the surface temperature, it reconstructs the depth-resolved temperature profile as heat propagates through the sample. As illustrated in Fig. 1d,e, two snapshots acquired at different times during heating exhibit distinct internal temperature distributions, resulting in different thermal-radiation spectra. By reconstructing the temperature profile at each time and tracking its temporal evolution, the method directly visualizes heat propagation within the material.

The temporal evolution of the reconstructed temperature field provides a direct measure of heat diffusion, enabling thermal conductivity to be determined without relying solely on indirect surface measurements. More broadly, transient depth thermography extends conventional depth thermography into a dynamic probe of heat transport, allowing thermal transport properties—including thermal diffusivity and interfacial thermal resistance—to be extracted from the measured spatiotemporal temperature field with substantially reduced dependence on model-based fitting of surface observables.

We experimentally validated the technique using two 3-mm-thick optical windows: fused silica and $MgF_2$ (both purchased from Thorlabs). The $MgF_2$ window is cut with the optical axis parallel to the c-axis of the crystal. We coated the backside of each window with a 100-nm-thick gold film to eliminate the influence of the non-ideal heater surface on the measured thermal-radiation spectra. Before each measurement, the heating stage (Linkam FTIR 600) was preheated to the desired temperature, and the room-temperature sample was gently placed in contact with the heater to initiate transient heat transfer.

Thermal radiation emitted from the top surface of the sample was collected in the normal direction using a 2-inch diameter gold-coated off-axis parabolic mirror (2-inch focal length) and directed into a Fourier-transform infrared (FTIR) spectrometer (Bruker Invenio-R). The FTIR was operated in repeated-scan mode, providing a temporal resolution of 0.126 s per spectrum (Supplementary Note 2). Figures 2a,b show representative thermal-radiation spectra acquired at t = 0.504, 1.764, and 3.024 s after the fused silica and $MgF_2$ windows were placed on a 200 °C heater.

The measured FTIR signal contains contributions from both the heated sample and the surrounding background [22]. To isolate the thermal radiation emitted by the sample, we calibrated our measurement setup using two reference samples with well-characterized optical properties, following the procedure described in Ref. [23] (Supplementary Note 3). The resulting calibrated thermal-radiation spectra are shown in Figs. 2c,d, where the purple- and grey-shaded regions denote the opaque and semi-transparent spectral ranges of the two windows used for further analysis. The pronounced dip near 9 μm in the fused-silica spectra arises from a phonon resonance [24].

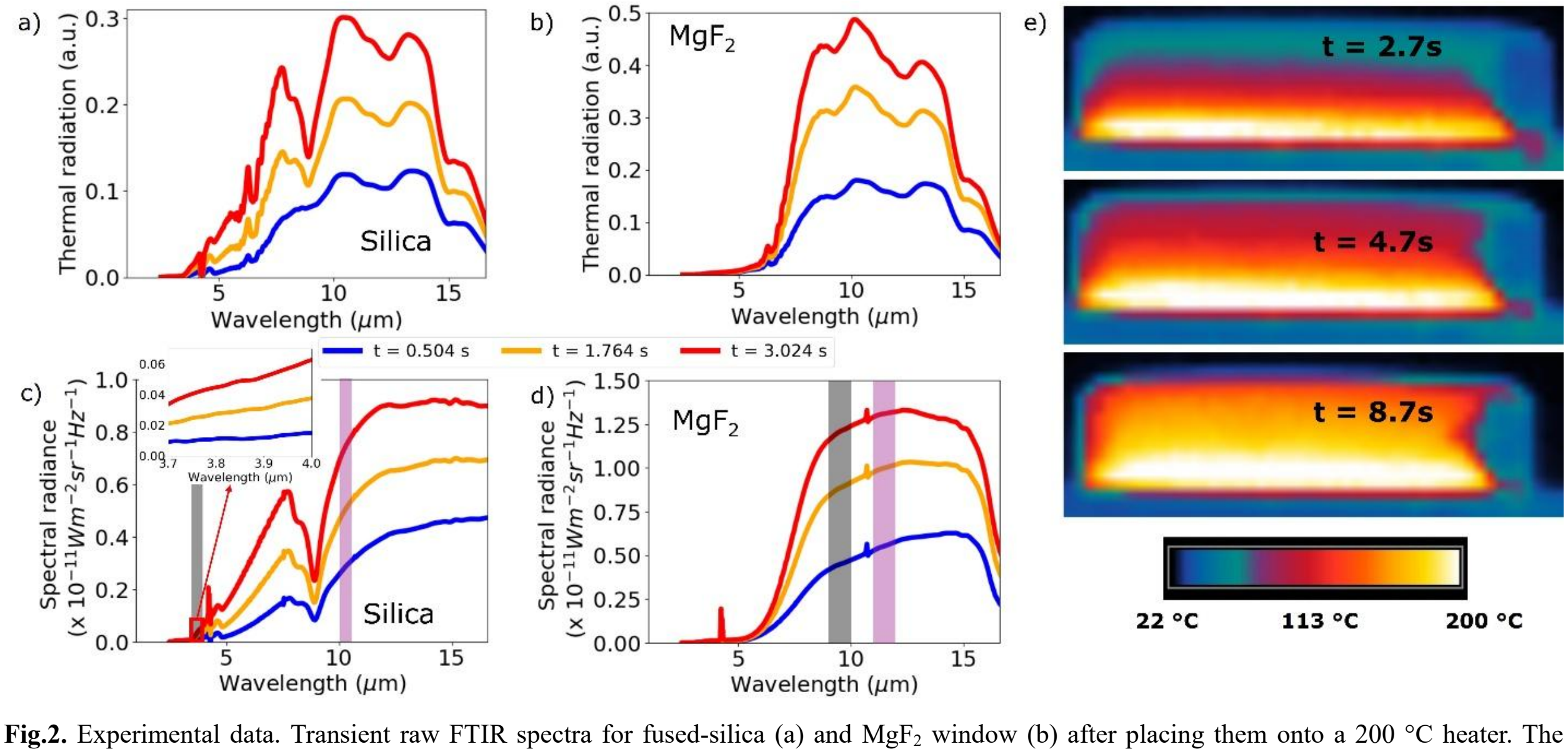

**Fig.2.** Experimental data. Transient raw FTIR spectra for fused-silica (a) and $MgF_2$ window (b) after placing them onto a 200 °C heater. The corresponding calibrated thermal-radiation spectra are plotted in (c) and (d). In (c) and (d), shaded magenta areas indicate the opaque ranges while shaded grey areas indicate the semitransparent ranges for each window used for data analysis. (e) Transient temperature profiles of the fused-silica window measured by an infrared camera at different times after the window was placed on the 200 °C heater.

Notably, the spectral radiance of the $MgF_2$ window in the opaque region is approximately 75% higher than that of the fused-silica window at the same time after heating. Because the two materials have nearly identical emissivities in their opaque spectral regions (Supplementary Note 1), this higher radiance directly indicates a higher surface temperature, reflecting more rapid heat diffusion through the $MgF_2$ window.

Because the thermal conductivity of most materials is temperature dependent, we repeated the measurements with the heater maintained at different temperatures to determine the temperature-dependent thermal conductivity of both windows. To validate the depth-resolved temperature profiles reconstructed from the thermal-radiation spectra measured at the top surface, we independently recorded the transient temperature distributions using an infrared camera (FLIR A400) viewing the sample from the side. Representative depth-resolved temperature profiles of the fused-silica window acquired at $t$ = 2.7, 4.7, and 8.7 s after contact with the 200 °C heater are shown in Fig. 2e.

For a material with a depth-dependent temperature profile $T(z)$, the thermal-radiation spectrum emitted from the top surface can be expressed as a depth-integrated contribution across the material [3,25]:

$$I(\lambda) = \int \bar{\epsilon}_i(\lambda, z) I_{BB}[\lambda, T(z)] dz, \tag{1}$$

where $\bar{\epsilon}_i(\lambda, z)$ is the local emissivity density and $I_{BB}$ is the blackbody radiation described by Planck's law [26]. In principle, Eq. (1) can be inverted to reconstruct the internal temperature profile, $T(z)$, from the measured thermal-radiation spectrum, $I(\lambda)$. However, this inverse problem is inherently ill-posed, with strong sensitivity to experimental noise and measurement uncertainties [4].

To achieve a robust and physically meaningful inversion, we employ two strategies. First, rather than independently retrieving the temperature at each depth, we parameterize the internal temperature profile using a compact functional representation:

$$T(z) = a + be^{z/c} \tag{2}$$

where $a$, $b$, and $c$ are fitting parameters. This choice is motivated by the experimental configuration: for transient heating with a fixed-temperature boundary condition at the bottom surface, the internal temperature profile of the window can be accurately approximated by an exponential function (Supplementary Note 4). This parameterization substantially reduces the dimensionality of the inverse problem and enhances the robustness of the reconstruction.

Second, we employ a two-band spectral fitting strategy. The opaque spectral region is first used to determine the surface temperature (parameter a), as thermal radiation in this regime originates exclusively from the top surface (z = 0). The semitransparent spectral region, where radiation contains information about the internal temperature distribution, is then used to extract the remaining parameters (b and c). The spectral bands are selected according to their optical penetration depths: regions with penetration depths <0.1 mm are considered opaque, whereas regions with penetration depths of 0.5–5 mm are chosen for depth-sensitive reconstruction. Accordingly, for fused silica, we use 10–10.5 µm as the opaque band and 3.7–4 µm as the semitransparent band; for $MgF_2$, the corresponding bands are 11–12 µm and 9–10 µm, respectively (see Supplementary Note 1 for the wavelength-dependent penetration depths measured for both samples).

Heat transport within the samples is governed by the one-dimensional diffusion equation [27]:

$$\frac{\partial T}{\partial t} = \alpha \frac{\partial^2 T}{\partial z^2}, \tag{3}$$

where $\alpha = k/\rho c_p$ is the thermal diffusivity, with $k$ the thermal conductivity, $\rho$ the density, and $c_p$ the specific heat capacity. Density and specific heat capacity values are taken from the literature for both fused silica and $MgF_2$ (Supplement Note 5). The boundary conditions are defined by a fixed temperature at the bottom surface and convective heat loss at the top surface:

$$k \frac{dT}{dz}\Big|_{z=0} = h\left(T_{surface} - T_{room}\right), \tag{4}$$

where $h$ is the effective convective heat transfer coefficient. The transient temperature evolution therefore depends on both $k$ and $h$, which can be extracted from the measured transient thermal-radiation spectra.

To determine these parameters, we employ a two-step procedure. First, we analyze the steady-state thermal-radiation spectra to assess the influence of convection. Under steady-state conditions, the temperature distribution across the window is linear, which we can obtain by fitting the thermal-radiation spectrum within the semitransparent spectral range. We then assess the contribution of convection from the temperature gradient between the bottom and top surfaces of the window. Note that the bottom-surface temperature does not necessarily equal the heater temperature because of the thermal resistance at the interface. We find that convection can be neglected for the $MgF_2$ sample because of its relatively small temperature gradient, whereas it has a non-negligible effect on the fused-silica sample (Supplementary Note 6).

The second step differs for fused silica and $MgF_2$. For $MgF_2$, the transient measurements are used to determine only the thermal conductivity, $k$. Specifically, two time instances $t_1$ and $t_2$ $(t_1 < t_2)$ are selected during the heating process. The spectrum at $t_1$, $I(\lambda, t_1)$, is inverted using Eqs. (1) and (2) to reconstruct the temperature profile $T(z, t_1)$. This profile is then propagated forward in time using Eq. (3), assuming negligible convection at the top surface and treating $k$ as a fitting parameter, to predict $T(z, t_2)$. The optimal value of $k$ is obtained by minimizing the difference between the measured spectrum $I(\lambda, t_2)$ and the spectrum calculated from Eq. (1) using the predicted temperature profile (see Supplementary Note 7 for details).

For fused silica, where convection contributes significantly to heat transfer, the transient data are used to simultaneously extract both $k$ and the convection coefficient $h$. The combination of $k$ and $h$ that minimizes the discrepancy between the measured and calculated spectra is identified as the fitted solution. In addition, we independently estimate the convection coefficient at different temperatures using a Nusselt-number-based approach [27] considering the sample geometry (Supplementary Note 8). The convection coefficients obtained from the two approaches show good agreement, further validating the robustness of transient depth thermography.

Note that, in our analysis, we assume the thermal conductivity $k$ remains constant across the sample thickness during a single heating experiment. For high heater temperature, the sample exhibits a very large thermal gradient initially. Because the thermal conductivity of both fused silica and $MgF_2$ exhibits strong temperature dependence, we therefore restrict our analysis to time intervals close to steady state, where the temperature variation across the sample is relatively small. This moderate temperature gradient ensures the validity of the constant-$k$ approximation.

As an example, Figs. 3(a,b) and (d,e) illustrate the data analysis procedure with a heater temperature of 200 °C. The initial spectra (blue curves in a and d) are first inverted to reconstruct the temperature profiles (blue curves in b and e), which are then propagated forward in time using Eq. (3) with the fitted thermal conductivity. The resulting temperature profiles (orange and green curves in b and e) generate calculated spectra (dotted curves in a and d) that agree closely with the experimental measurements, validating the extracted thermal conductivity $k$. Independent infrared camera measurements of the depth-resolved temperature profiles (dotted curves in Figs. b and e) further confirm the accuracy of the reconstructed temperature distributions from transient depth thermography.

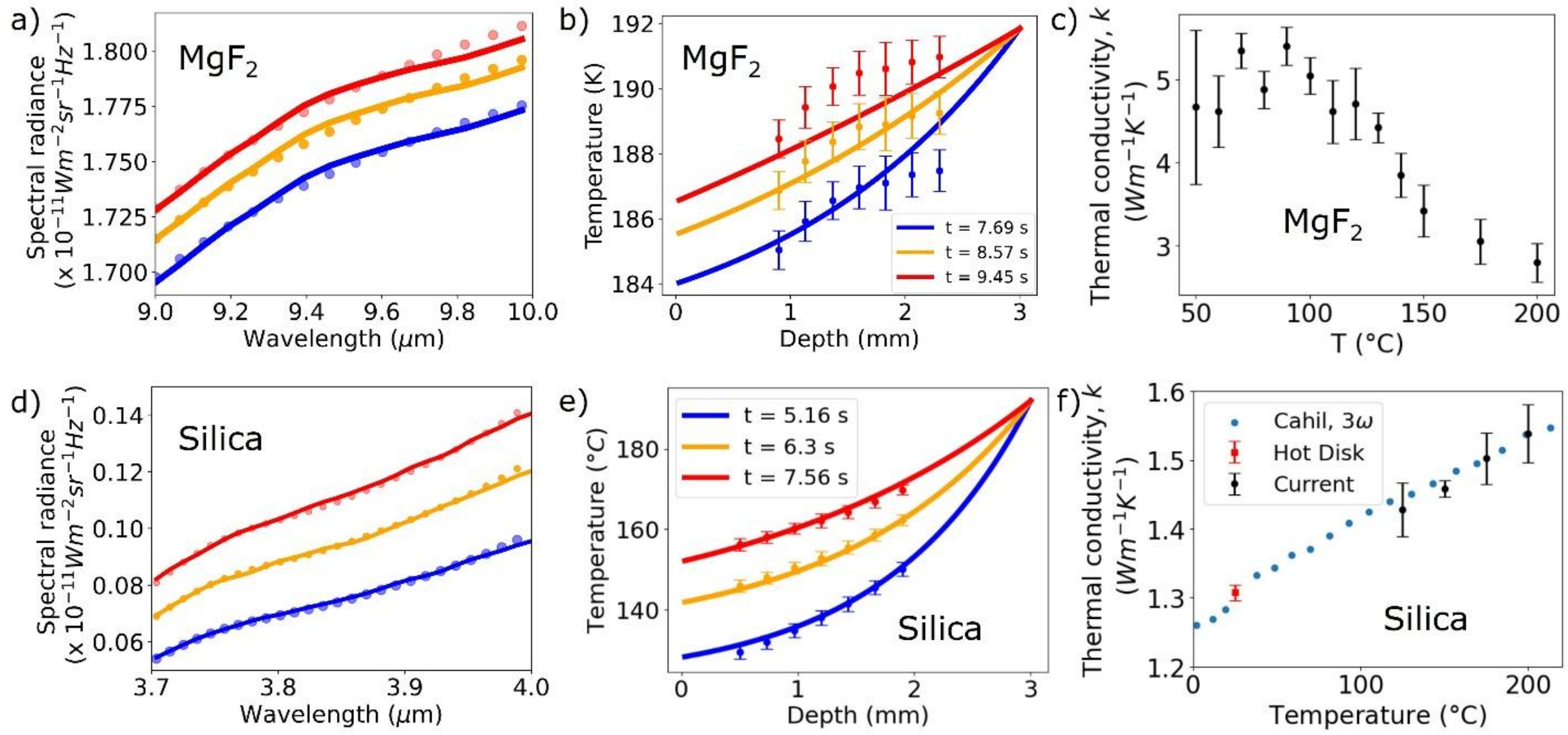

**Fig.3.** Thermal conductivity fitting for $MgF_2$ (a-c) and fused silica (d-f) windows. (a, d) Measured (solid) and calculated (dotted) transient thermal-radiation spectra for windows placed onto a 200°C heater. (b, e) Corresponding transient temperature profiles, obtained via transient depth thermography (solid) and infrared camera measurements (dotted). (c) and (f) Temperature-dependent thermal conductivity of $MgF_2$ and fused silica. In (f), literature data from 3ω method [8] are shown by dot and room temperature value of our fused-silica window measured by the hot disk method is shown by square.

By repeating this procedure at different heater temperatures, we extract the temperature-dependent thermal conductivity of both materials. For fused silica, measurements were performed at 125, 150, 175, and 200 °C, with the temperature lower bound limited by its relatively weak thermal radiation in the semitransparent spectral region (~4 μm). In contrast, $MgF_2$ exhibits stronger emission near its semitransparent region (~9 μm), which overlaps with the peak of blackbody radiation, enabling measurements at a much lower temperature. Specifically, measurements were conducted from 50 to 150 °C in 10 °C increments, along with additional measurements at 175 and 200 °C. The extracted thermal conductivities are summarized in Figs. 3c and 3f. Each data point represents the average of more than three independent measurements, with approximately 200 combinations of $t_1$ and $t_2$ analyzed for each experiment. Error bars denote the standard error.

For validation, the measured thermal conductivity of fused silica is compared with literature values obtained using the 3ω method (dotted, Fig. 3f) [8]. The results show excellent agreement and accurately capture the expected temperature dependence. We further measured the thermal conductivity of our fused silica window using the transient plane source (hot disk) method at room temperature (Supplementary Note 9), obtaining consistent results with the 3ω measurement.

In contrast, the thermal conductivity of $MgF_2$ is significantly less documented, particularly at elevated temperatures. Even at room temperature, reported values vary substantially among manufacturers, including 0.3 [28], 11.6 [29], and 21 W $m^{-1}$ $K^{-1}$ [30]. To the best of our knowledge, only one early study has reported measurements of $MgF_2$ thermal conductivity, and it was limited to cryogenic temperatures below 77 K [31]. Our measurements reveal that the thermal conductivity of $MgF_2$ remains approximately constant between 50 and 100 °C, followed by a gradual decrease at higher temperatures.

Although demonstrated here using millimeter-scale bulk samples, transient depth thermography can be extended to thin films with buried interfaces. The primary requirement is matching the temporal resolution of the measurement to the faster thermal dynamics associated with reduced length scales. This can be achieved by replacing the steady-state heater with a pulsed excitation source, such as a laser, and employing high-speed spectroscopic detection. Modern FTIR systems operated in step-scan mode can achieve nanosecond temporal resolution, providing a viable route for probing rapid thermal processes [32]. These advances will enable the application of transient depth thermography to nanoscale and multilayer materials, where accurate depth-resolved characterization of thermal transport remains a major challenge.

As a numerical example, we consider a two-layer sample consisting of a 500-nm-thick GaN layer on top of a 500-nm-thick AlGaN layer, representative of structures used in GaN RF devices [33,34] (Fig. 4a). The sample is heated by a CW laser with a flux of 700 MW/$m^2$ and we consider the transient heating immediately after laser heating. We consider a range of thermal boundary resistance (TBR) values, spanning from well-fabricated interfaces [35] to progressively poorer interfaces characterized by higher TBRs. Solving the heat-transfer equation on this two-layer structure (Supplementary Note 11), we pick three time instances at which the temperature profiles in the GaN layer are nearly identical but differ dramatically within the AlGaN layer (Fig. 4(b)). This difference cannot be directly captured by conventional surface-sensitive techniques. In contrast, the thermal-radiation spectra (Fig. 4(c)) clearly reveal that the temperature of the AlGaN layer decreases as TBR increases, further allowing the determination of TBR value. This capability illustrates a major advantage of transient depth thermography over existing techniques, in which the thermal boundary resistance is commonly inferred indirectly by fitting thermal-transport models to surface-measured signals [11,21,36].

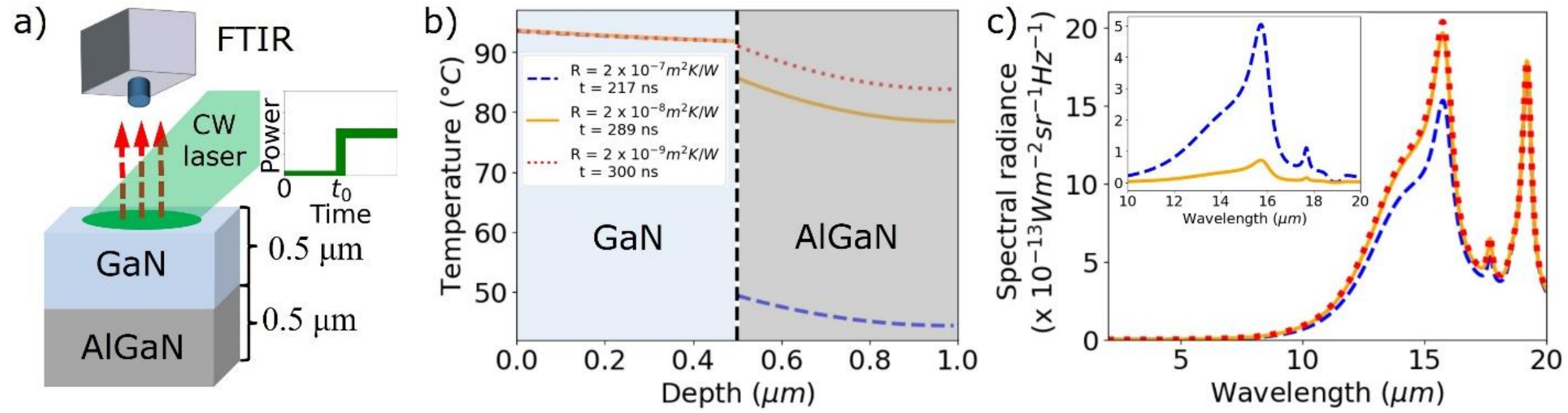

**Fig.4.** Transient depth thermography reveals thermal boundary resistance. (a) A room-temperature sample (a stacked GaN/AlGaN structure is shown as an example) is heated by a CW laser, and its transient thermal-radiation spectra immediately after the laser heating are measured using an IR spectrometer. (b) Simulated temperature profiles corresponding to nearly identical temperatures in the GaN layer for different values of thermal boundary resistance (TBR). (c) The corresponding thermal-radiation spectra. Inset shows the spectral differences between the spectrum obtained with the lowest TBR and those obtained with higher TBR (blue solid line) and the highest TBR (red dashed line).

## Discussion

A comparison of transient depth thermography with established thermal metrology techniques is summarized in Table I. Transient depth thermography distinguishes itself by directly measuring both surface and depth-resolved temperatures through time-resolved thermal-radiation spectra. This depth sensitivity provides direct access to the spatiotemporal evolution of heat transport within a material, reducing reliance on indirect observables and complex model-based fitting while achieving an uncertainty of 2–7%. Unlike conventional surface-sensitive techniques, including thermoreflectance, 3ω, and laser flash methods, our approach directly captures internal temperature dynamics and is applicable to both bulk materials and thin-film structures. Compared with steady-state methods, it is non-destructive and avoids the stringent requirements for well-defined boundary conditions, enabling measurements of thin films and complex multilayer systems. By combining depth- and time-resolved temperature measurements, transient depth thermography provides a physically transparent and quantitatively accurate framework for characterizing thermal conductivity and heat transport in complex materials.

The applicability of transient depth thermography relies on the presence of a spectral window where the optical penetration depth is comparable to the sample thickness. This condition is generally not met in opaque materials, such as bulk metals and highly doped semiconductors, where limited optical penetration prevents depth-sensitive measurements. In addition, accurate temperature reconstruction requires sufficient signal-to-noise ratio (SNR) within the relevant spectral range. For example, the optimal spectral windows are located near 10 μm for $MgF_2$ and 4 μm for fused silica. Because blackbody radiation from near-ambient-temperature objects peaks near 10 μm, $MgF_2$ can be characterized at relatively low temperatures (down to ~50 °C), whereas fused silica requires elevated temperatures (>100 °C) to generate sufficient emission near its semitransparent spectral window. This limitation is not intrinsic to the technique and can be mitigated through advances in detector sensitivity and infrared detection technologies.

**Table 1. Comparison of thermal conductivity measurement techniques.**

| Method | Applicable Samples | Temperature Information | Measurement Principle | Typical Uncertainty |
|---|---|---|---|---|
| Thermoreflectance | Thin films | Surface temperature | Indirect: optical reflectance combined with thermal modeling | 10–100% [12]* |
| 3ω Method | Thin films | Surface temperature | Indirect: electrical response combined with thermal modeling | up to 50% [37] or even higher [38]* |
| Laser Flash | Bulk materials and thin films | Surface temperature | Direct: surface temperature response following pulsed heating | 3-5% (bulk) [39] |
| Hot Disk | Bulk materials | Surface temperature | Direct: surface temperature measurement using thermocouples | ~5% [40] |
| Steady-State Methods | Bulk materials | Surface and depth temperature | Direct: temperature-gradient measurement using thermocouples | 2-20% [39] |
| **Transient Depth Thermography** | Bulk materials and thin films | Surface and depth temperature | Direct: thermal-radiation spectra with transient temperature reconstruction | 2–7% (bulk)** |

*For thin films uncertainty usually increases with decrease in thickness.

**We obtained 2% uncertainty for fused-silica's thermal conductivity and 7% for $MgF_2$.

Large temperature gradients introduce additional considerations in the analysis. When the temperature variation across the sample becomes significant, the assumption of a spatially uniform thermal conductivity may no longer be valid. Under these conditions, the extracted effective thermal conductivity can depend on the selected time interval, particularly for materials with strong temperature-dependent thermal properties (Supplementary Note 10). At the same time, this regime presents a unique opportunity: the rich spatiotemporal information encoded in thermal-radiation spectra may enable the direct extraction of temperature-dependent thermal conductivity from a single transient measurement, eliminating the need for multiple experiments at different heater temperatures. Although the present work focuses on a constant-$k$ analysis framework, this capability highlights the broader potential of transient depth thermography for advanced thermal characterization under nonuniform temperature conditions.

The current implementation of transient depth thermography offers several opportunities for further improving the measurement accuracy. First, incorporating temperature-dependent density and heat capacity would improve the determination of temperature-dependent thermal conductivity, since the method primarily measures thermal diffusivity, as do other transient techniques such as the laser flash method [15]. Second, adopting more uniform temporal sampling during transient measurements could further enhance accuracy. For example, the step-scan mode of an FTIR spectrometer enables repeated measurements with nanosecond-to-microsecond time resolution. Finally, reducing thermal gradients during heating would minimize errors associated with large initial temperature gradients when the sample is placed on the heated stage. This limitation could be overcome by replacing stage heating with pulsed laser heating.

Looking forward, transient depth thermography establishes a new paradigm for thermal metrology by integrating spectral, spatial, and temporal information into a unified measurement framework. This capability enables high-precision characterization of heat transport in complex material systems, including nanostructures and multilayer architectures, with broad implications for fundamental studies of thermal physics and the development of next-generation photonic and electronic devices.

## Materials and methods

The samples are half-inch diameter fused silica and MgF2 windows of the thickness of 3 mm from Thorlabs. The heater is Linkam FTIR 600 with a temperature stability of 0.01 °C. The Fourier-transform infrared (FTIR) spectrometer is Bruker Invenio-R equipped with a nitrogen-cooled HgCdTe (MCT) detector. The infrared camera is a FLIR A400. The ellipsometer is J.A. Woollam IR-VASE Mark II. More details of the experiments and modeling can be found in the Supplementary.

**Acknowledgements**
D.S., C. W. , and Y. X acknowledge the support from DARPA YFA program (Grant No. D23AP0018900) and NSF LEAPS-MPS program (Award Number 2418002).

**Author contributions**
D.S. and Y.X. conceived of the project and designed the experiments and simulations. D.S. and Y.X. built the experimental setup. D.S., C.W., Y.J, and R.S. conducted the experiments. D.S. performed numerical simulations. All of the authors discussed the results and contributed to the writing of the manuscript. Y.X. and N.D'S. supervised the project.

**Conflict of interest**
The authors declare that they have no conflict of interest.

# Supplementary information

# Transient Depth Thermography for Probing Heat Transport

Dmitrii Shymkiv[1], Chun Wang[1], Yan Jiang[1], Rajat Srivastava[2], Nandika D'Souza[2], and Yuzhe Xiao[1*]

*[1]Department of Physics, University of North Texas, Denton, TX 76203, USA*
*[1]Department of Mechanical Engineering, University of Texas at Dallas, Richardson, TX 75080, USA*
**Email address: yuzhe.xiao@unt.edu*

**Supplement 1. Optical characterization of 3-mm fused-silica and $MgF_2$ windows.**

We performed optical characterization of the samples using a J.A. Woollam IR-VASE Mark II ellipsometer. The ellipsometric parameters, Ψ and Δ, were measured at two incidence angles (65° and 75°) and fitted using Gaussian oscillator models to extract the real (**n**) and imaginary (**k**) components of the refractive index. In addition, transmittance measurements were performed to determine the imaginary component of the refractive index in the semitransparent spectral region.

To account for the temperature dependence of the optical properties, transmittance spectra were acquired at multiple temperatures. In the semitransparent wavelength range, we assume that only the imaginary part of the refractive index exhibits significant temperature dependence due to its relatively large variation, while temperature-induced changes in the real part of the refractive index are neglected because of their relative small magnitude. The measured and fitted optical properties are summarized in Fig. S1.1 for fused silica and Fig. S1.2 for $MgF_2$.

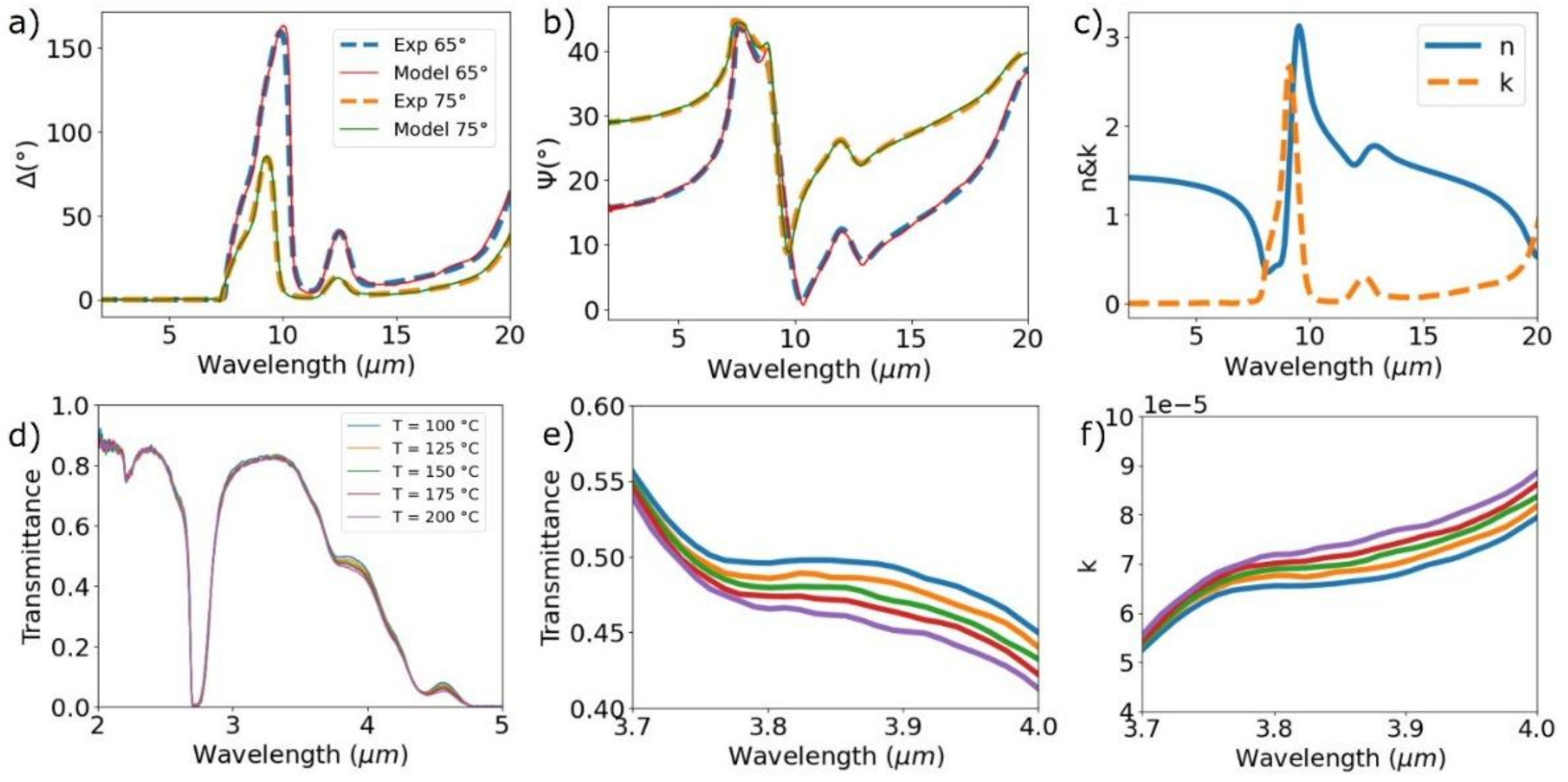

**Fig. S1.1.** Optical characterization of a fused-silica window. Measured ellipsometric parameters: (a) Δ and (b) Ψ obtained using IR-VASE. (c) Extracted real (n) and imaginary (k) components of the refractive index from ellipsometric fitting. (d) Temperature-dependent transmittance spectra measured through the fused-silica window and (e) enlarged view of the spectral region of interest. (f) Extracted temperature-dependent imaginary component (k) of the refractive index.

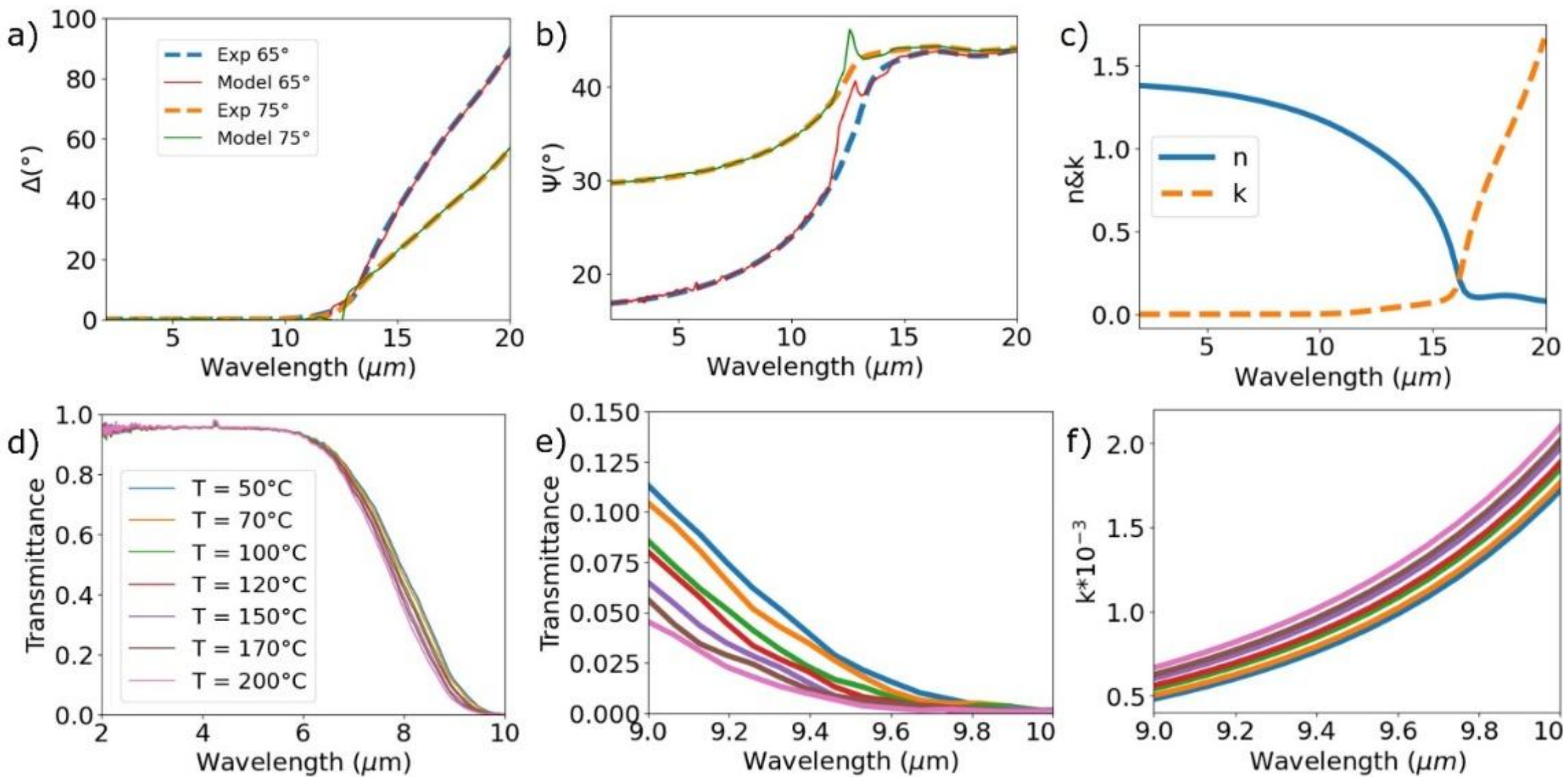

**Fig. S1.2.** Optical characterization of a $MgF_2$ window. Measured ellipsometric parameters: (a) Δ and (b) Ψ obtained using IR-VASE. (c) Extracted real (n) and imaginary (k) components of the refractive index from ellipsometric fitting. (d) Temperature-dependent transmittance spectra measured through the $MgF_2$ window and (e) enlarged view of the spectral region of interest. (f) Extracted temperature-dependent imaginary component (k) of the refractive index**.**

Once the imaginary part of the refractive index ($k$) is known, the penetration depth can be calculated by $\delta = \frac{\lambda}{4\,\pi\,k}$ (Fig.S1.3).

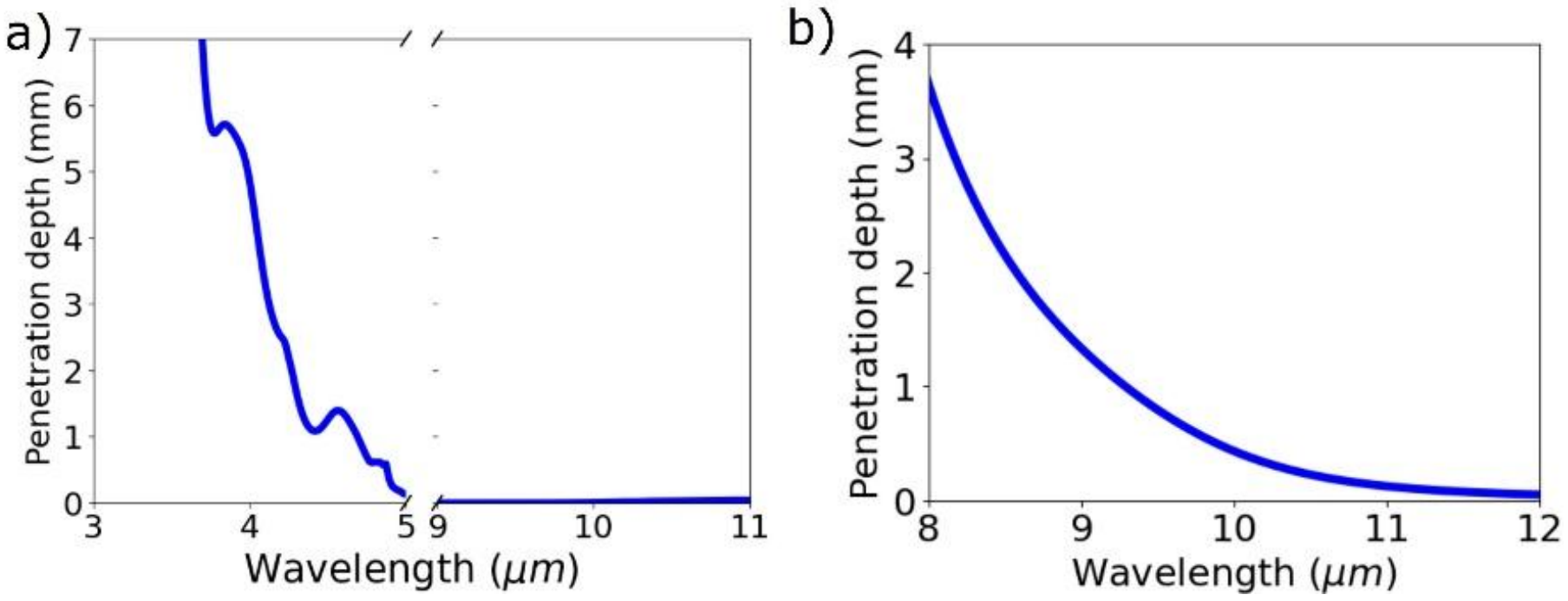

**Fig.S1.3** Wavelength-dependent penetration depth for (a) fused silica and for (b) $MgF_2$.

The primary purpose of the optical characterization in this work is to determine the real and imaginary components of the refractive index, which enable calculation of the local emissivity within the samples. These optical properties allow us to transform depth-resolved temperature profiles into corresponding thermal-radiation spectra and provide the necessary foundation for solving the inverse problem in transient depth thermography. Figure S1.4 shows the calculated total emissivities of 3-mm-thick fused-silica and $MgF_2$ samples.

Note that the emissivity in the semitransparent spectral range can exceed 1-transmittance due to the presence of the metal coating on the backside of the samples. In the opaque spectral range, the emissivities of fused silica and $MgF_2$ are comparable. Because the optical penetration depth is small in this wavelength range, the measured emissivity primarily represents the local surface emissivity.

As shown in Fig. 2c,d, the thermal-radiation spectra provide qualitative evidence that $MgF_2$ exhibits faster thermal diffusion than fused silica. The spectra shown for the two samples correspond to identical time points after heating, yet their absolute radiance values differ substantially. Since the emissivities in the opaque region are comparable, the higher thermal radiance observed for $MgF_2$ indicates a higher surface temperature, suggesting more rapid heat propagation through the sample.

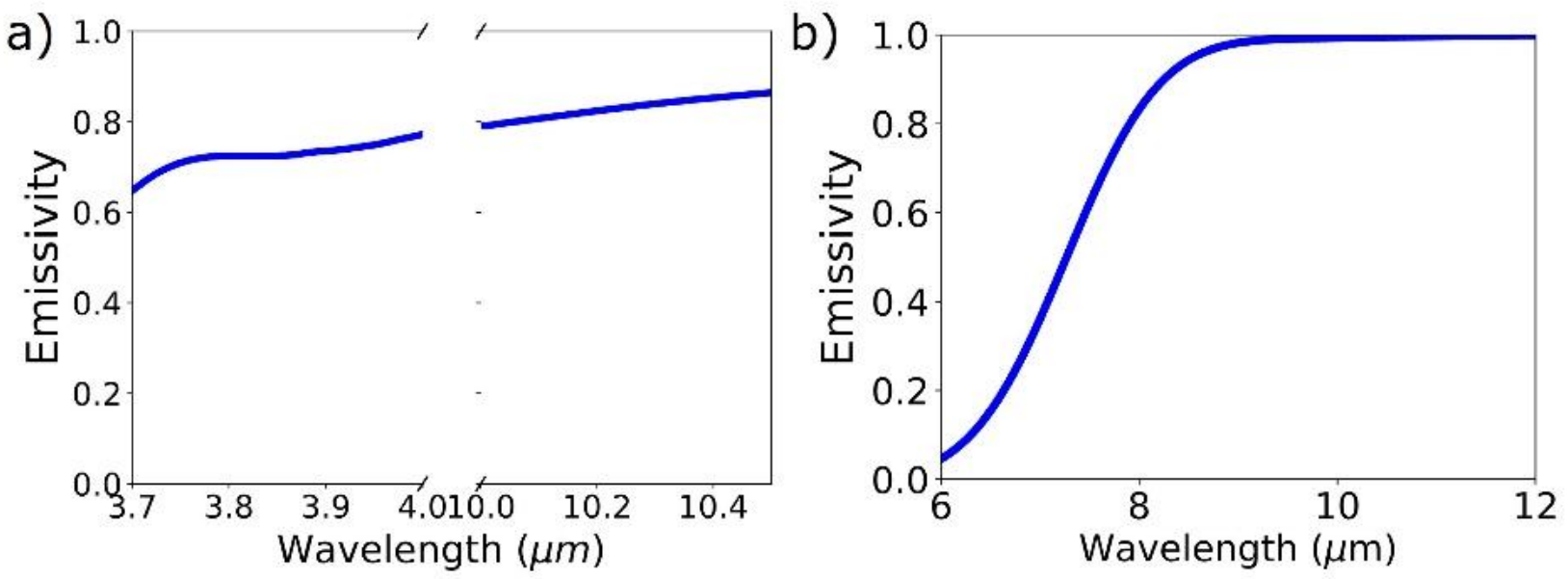

**Fig.S1.4** Wavelength-dependent total emissivity calculated for (a) fused silica and for (b) $MgF_2$.

However, the temperature gradients within the samples remain large during this transient heating process, with the bottom surface near 200 °C and the top surface only on the order of 50 °C. Therefore, quantitative conclusions cannot be drawn solely from the spectra in Fig. 2c,d, where the samples are far from thermal equilibrium. Instead, a rigorous comparison of thermal diffusivity requires the full transient analysis presented in Supplementary Note 10, which confirms that $MgF_2$ exhibits higher average thermal diffusivity than fused silica.

## Supplement 2. Determine the time resolution of the FTIR

We determined the temporal resolution of the FTIR repeated-scan mode using the following procedure. First, we measured the temperature-dependent thermal radiation of a heated sapphire sample, which has a relatively high thermal conductivity (~40 W $m^{-1}$ $K^{-1}$). Figure S2a shows the thermal-radiation intensity at 10.17 µm, corresponding to the opaque spectral region of sapphire.

We then measured the transient thermal radiation from the sapphire placed on the heater using the repeated-scan mode. The heater was initially at room temperature, and the maximum available heating rate of 2.5 °C $s^{-1}$ was applied (Fig. S2b). The temporal step between consecutive FTIR spectra was determined by correlating the measured temperature change between two consecutive spectra with the known heating rate (Fig. S2c).

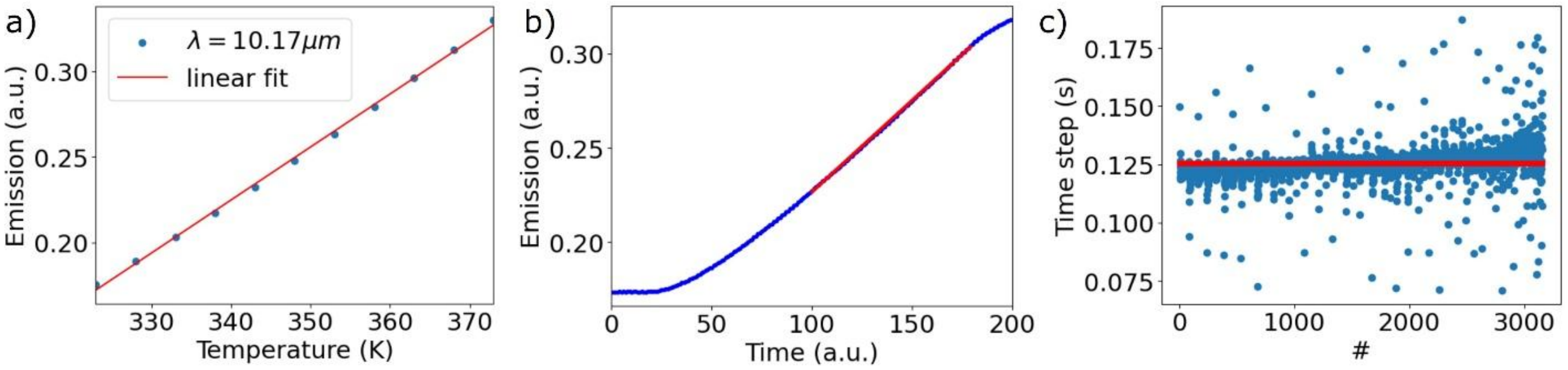

**Fig. S2.** Determination of the temporal step in the FTIR repeated-scan measurement mode. (a) Temperature-dependent thermal-emission signal at $\lambda$ = 10.17 µm measured from a heated sapphire wafer at different temperatures. The red solid line represents a linear fit to the data. (b) Transient thermal-emission signal at $\lambda$ = 10.17 µm measured from a sapphire wafer placed on the heater stage using the repeated-scan mode. The heater temperature was increased at a rate of 2.5 °C $s^{-1}$. The data points highlighted by the red solid line indicate the measurements used for temporal-step determination. (c) Calculated temporal steps obtained from different pairs of data points within a single repeated-scan measurement sequence. The red line indicates the average temporal step of 0.126 s.

## Supplement 3. Thermal-radiation measurement system calibration.

We used an FTIR to measure thermal radiation from the samples. The measured FTIR signal contains contributions from both the heated sample and the surrounding background [S1]. To isolate the thermal radiation emitted by the sample, we calibrated the measurement system using two reference samples with well-characterized optical properties, following the procedure described in Ref. S2. More specifically, the FTIR measured signal from the emitter x can be written as:

$$S_x(\lambda) = m(\lambda)[I_x(\lambda) + B_1(\lambda) + R_x(\lambda)B_2(\lambda)], \quad \text{(S3.1)}$$

where $m(\lambda)$ is the system response, $B_1(\lambda)$ and $R_x(\lambda)B_2(\lambda)$ are the sample-independent and sample-dependent background emission, $R_x(\lambda)$ is the reflectivity of the emitter, and $I_x(\lambda)$ is the true emission spectrum from the emitter.

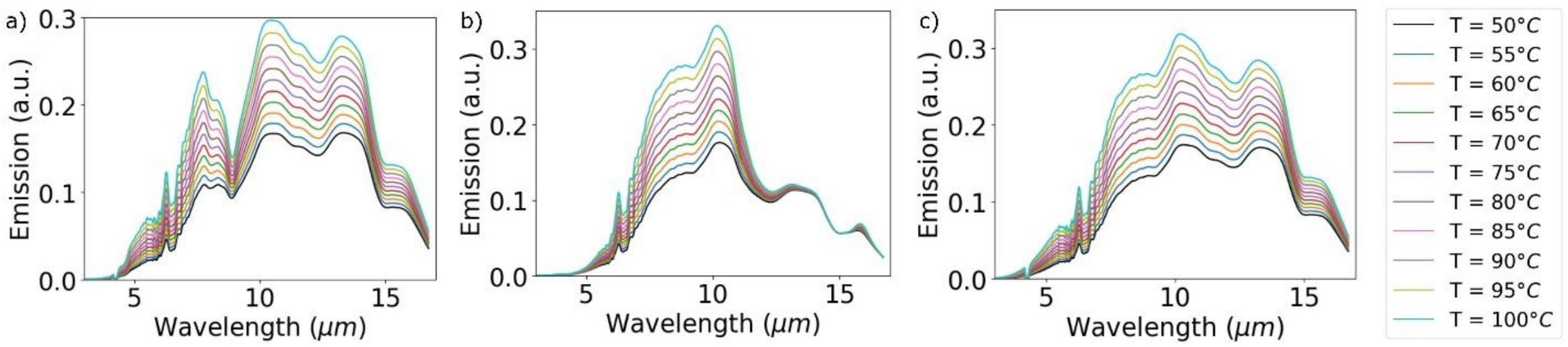

**Fig.S3.1** Measured thermal-radiation spectra from (a) a 0.5-mm fused silica wafer, (b) a 0.5-mm sapphire wafer, and (c) a carbon nanotube forest (CNT) blackbody reference at temperatures from 50 to 100 °C in 5 °C increments.

If the emitter is uniformly heated, then true emission spectrum is expressed as:

$$I_x(\lambda, T) = \epsilon_x(\lambda, T) I_{BB}(\lambda, T), \qquad \text{(S3.2)}$$

where $\epsilon_x(\lambda, T)$ is the emissivity of the emitter and $I_{BB}(\lambda, T)$ is the blackbody radiation given by the Planck's law.

$$\epsilon(\lambda) = \epsilon_{BB}(\lambda) \frac{S(\lambda,T_2) - S(\lambda,T_1)}{S_{BB}(\lambda,T_2) - S_{BB}(\lambda,T_1)} \qquad \text{(S3.3)}$$

Once the emissivity of the sample is known, system response function can be calculated as:

$$m(\lambda) = \frac{S_x(\lambda,T_2) - S_x(\lambda,T_1)}{\epsilon_x(\lambda)[I_{BB}(\lambda,T_2) - I_{BB}(\lambda,T_1)]} \qquad \text{(S3.4)}$$

To calibrate our FTIR spectrometer we measured thermal-radiation spectra from three reference samples - a 0.5-mm fused silica wafer, a 0.5-mm sapphire wafer, and a carbon nanotube forest (CNT) blackbody reference at temperatures from 50 to 100 °C in 5 °C increments (Fig. S3.1). Then we determined emissivity and system response from this data (Fig. S3.2).

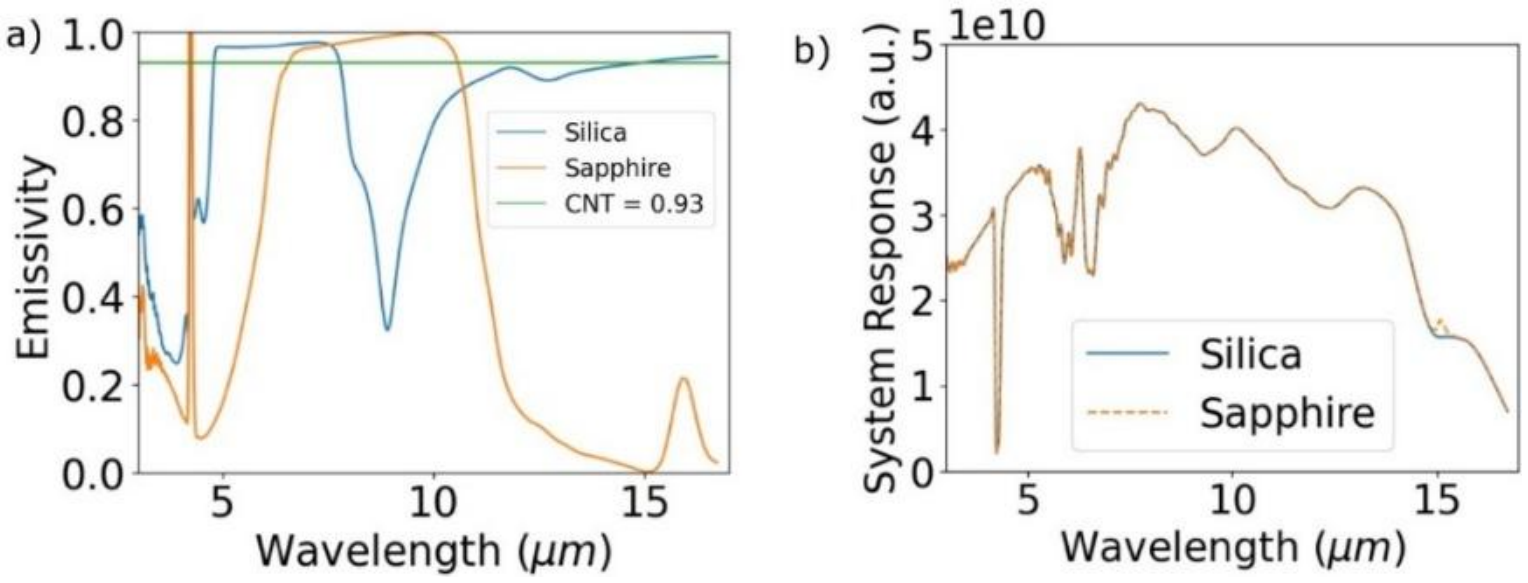

**Fig.S3.2** (a) Measured emissivity of the reference samples. (b) Calibrated system response function.

The next step in the calibration process is to determine background components $B_1$ and $B_2$ (Fig. S3.3). The total background emission from a sample *x* is given by:

$$B_x(\lambda) = B_1(\lambda) + R_x(\lambda) B_2(\lambda) = S_x(\lambda, T_1)/m(\lambda) - \epsilon_x(\lambda) I_{BB}(\lambda, T_1) \qquad \text{(S3.5)}$$

To calculate individually $B_1$ and $B_2$, at least two different samples (*x* and *y*) are required:

$$B_2(\lambda) = \frac{B_x(\lambda) - B_y(\lambda)}{R_x(\lambda) - R_y(\lambda)}, \qquad B_1(\lambda) = B_x(\lambda) - R_x(\lambda) B_2(\lambda) \qquad \text{(S3.6)}$$

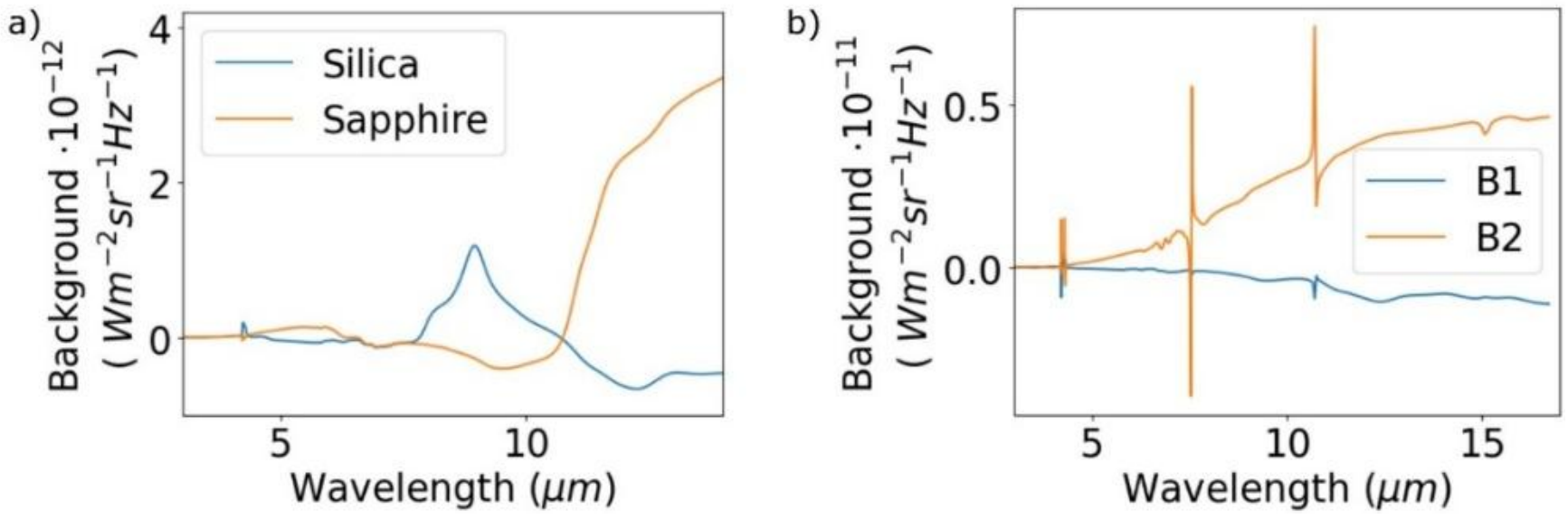

**Fig.S3.3** (a) Total background emission for fused silica and sapphire reference samples. (b) Calibrated sample-independent $B_1$ and $B_2$.

**Supplement 4. Approximate the sample temperature with an exponential function.**
We simulated the experiment by solving the heat diffusion equation with a prescribed thermal diffusivity. The resulting temperature profiles at all time points were accurately described by exponential functions, yielding an average fitting error of 0.5%. Figure S4 shows representative temperature profiles corresponding to the best and worst fitting cases. These results validate the use of exponential functions to approximate the temperature distributions within the samples in our analysis.

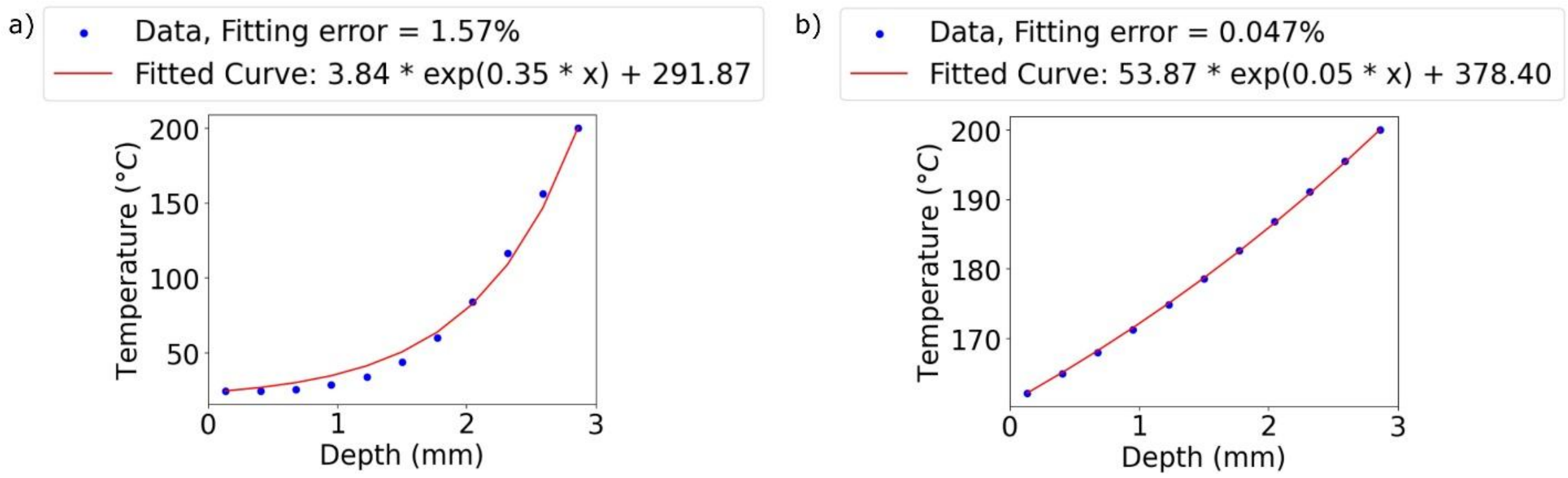

**Fig.S4** Exponential fitting of temperature profiles obtained from transient heating simulations. The figures show representative temperature profiles corresponding to the largest (a) and smallest (b) fitting errors for the exponential approximation.

**Supplement 5. Density and heat capacity for fused silica and $MgF_2$.**
The following values of density and heat capacity for fused silica and MgF2 are used in this work [S3,S4].

**Table S5.1 Density and heat capacity for fused silica and $MgF_2$.**

| Material | Density (kg/m$^3$) | Heat capacity (J/kg K) |
|---|---|---|
| Fused silica | 2200 | 740 |
| $MgF_2$ | 3180 | 1000 |

**Supplement 6. Steady-state measurements analysis.**
In addition to the transient thermal-radiation measurements, we acquired steady-state thermal-radiation spectra from samples placed on the heater at different temperatures. The surface temperature was first determined from the opaque spectral region. Assuming a linear temperature distribution across the sample thickness under steady-state conditions, the depth-resolved temperature profiles were then reconstructed through spectral inversion. Importantly, the bottom temperature of the sample was not constrained to equal the heater temperature, allowing for possible thermal contact resistance and nonideal thermal coupling between the sample and heater to be considered. The reconstructed temperature profiles for fused silica and $MgF_2$ with heater temperatures of 100 and 150 °C are summarized in Figs. S6.1 and S6.2, respectively.

For fused silica, the temperature gradient across the sample increases from 4 to 20 °C as the heater temperature rises from 100 to 200 °C, indicating that convection has a significant influence on the temperature profiles. In contrast, for $MgF_2$, the maximum temperature gradient remains below 4 °C even when the heater temperature reaches 200 °C. Therefore, the effect of convection can be neglected for $MgF_2$ within the investigated temperature range.

By analyzing the steady-state data, we observe that the bottom temperature of the sample can differ substantially from the heater temperature in some experiments (Fig. S6.2e, f). The dash-dotted black line in Fig. S6.2e represents the calculated thermal-radiation spectrum obtained using a linear temperature profile with the bottom temperature constrained to equal the heater temperature. This calculated spectrum exceeds the measured spectrum, particularly at shorter wavelengths where emission from deeper regions of the window contributes more strongly. This discrepancy indicates that the bottom region of the sample is at a lower temperature than the heater, resulting in a reconstructed temperature profile (Fig. S6.2f) with a bottom temperature more than 3 °C below the heater temperature.

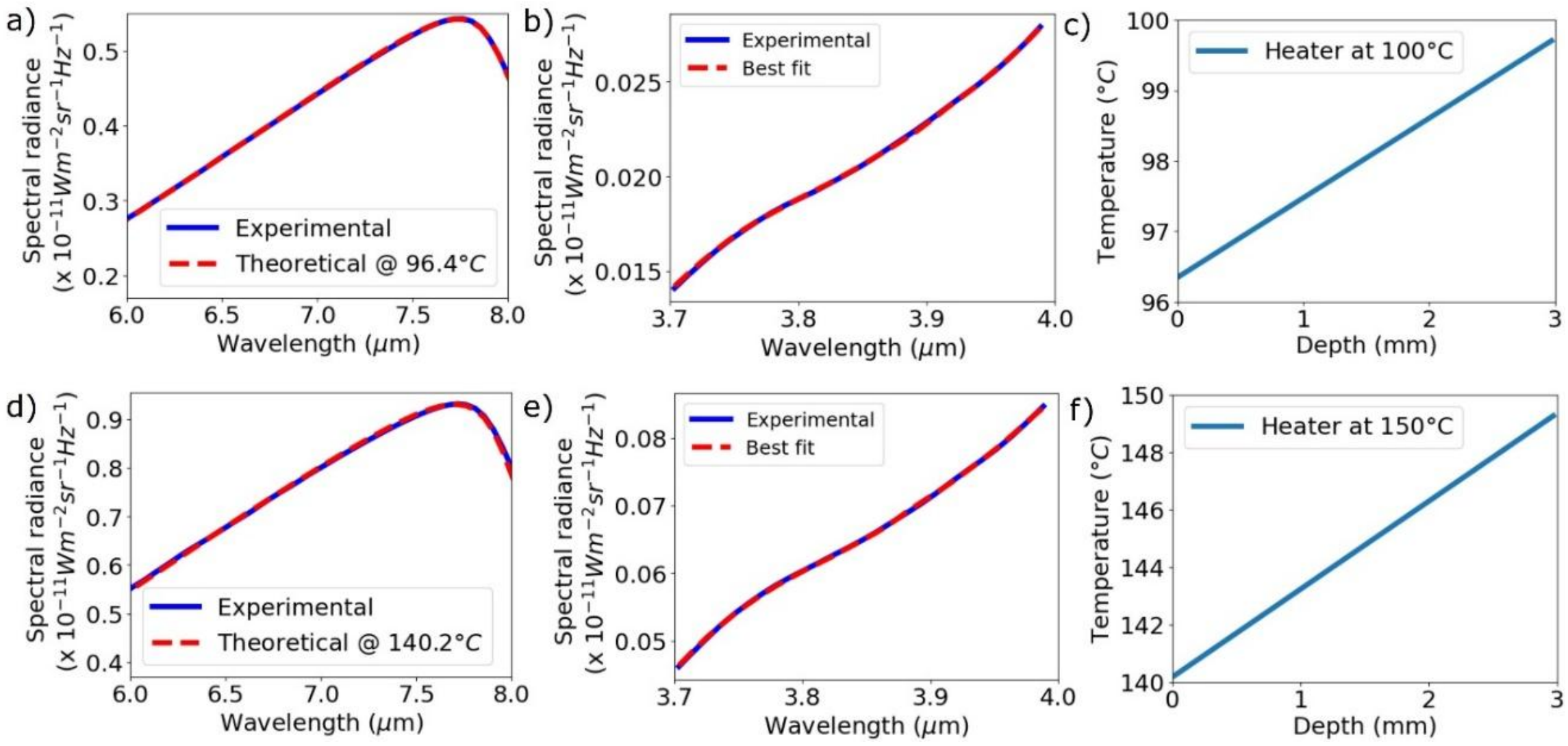

**Fig.S6.1** Steady-state measurements of a fused-silica window on the heater at 100 °C (a–c) and 150 °C (d–f). (a,d) Measured (solid curves) and calculated (dashed curves) thermal-radiation spectra in the opaque spectral region using the reconstructed temperature profiles. (b,e) Measured and calculated thermal-radiation spectra in the semitransparent spectral region. (c,f) Reconstructed depth-resolved temperature profiles obtained from spectral inversion.

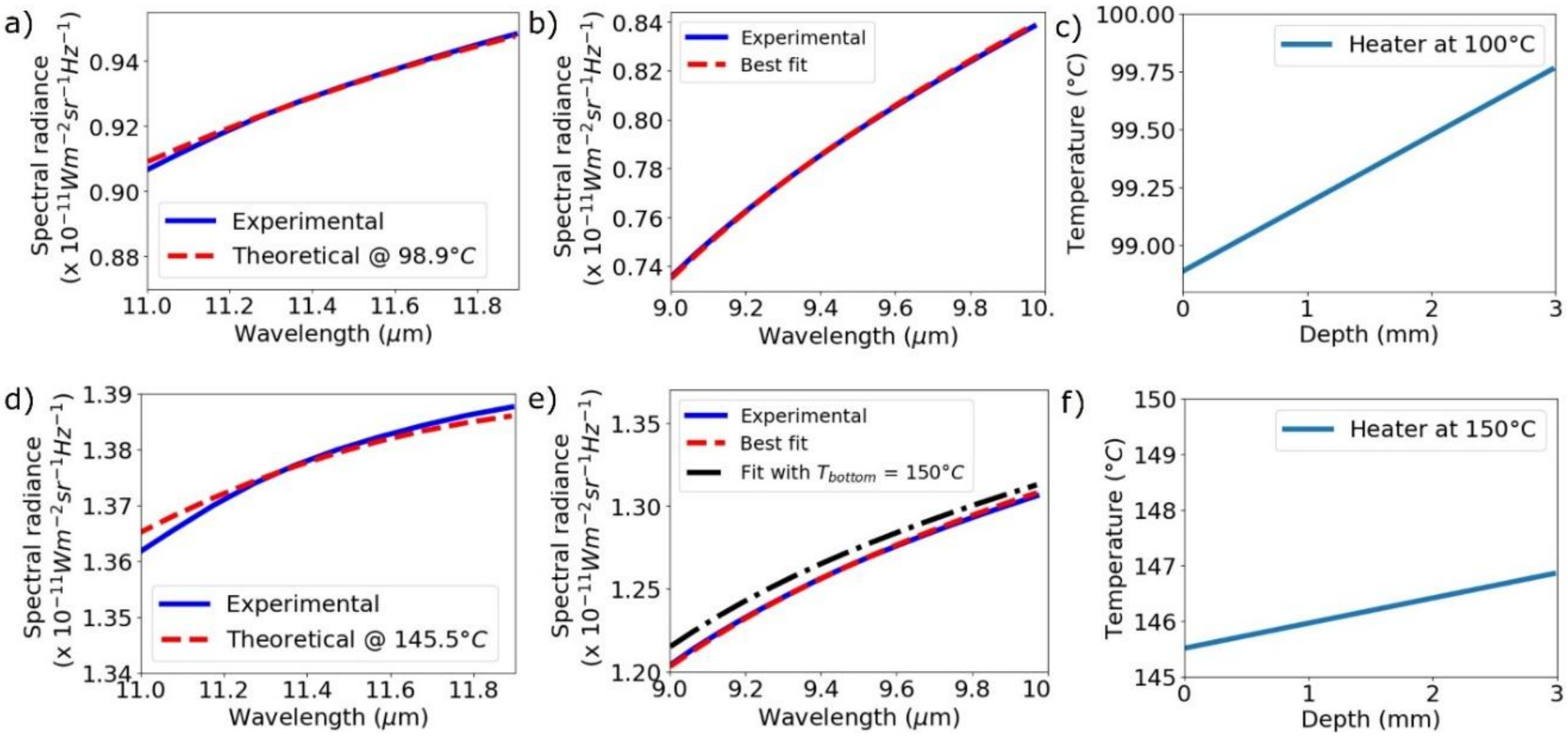

**Fig.S6.2** Steady-state measurements of an $MgF_2$ window on the heater at 100 °C (a–c) and 150 °C (d–f). (a,d) Measured (solid curves) and calculated (dashed curves) thermal-radiation spectra in the opaque spectral region using the reconstructed temperature profiles. (b,e) Measured and calculated thermal-radiation spectra in the semitransparent spectral region. (c,f) Reconstructed depth-resolved temperature profiles obtained from spectral inversion. Note that the bottom temperature of the $MgF_2$ window is significantly lower than the heater temperature, indicating the presence of thermal contact resistance between the sample and heater.

## Supplement 7. Determination of thermal conductivity.

In this Supplementary Note, we provide a detailed description of the procedure used to determine thermal conductivity. A single $MgF_2$ measurement with the heater temperature set to 70 °C is presented as an example. All other measurements and heater temperatures were analyzed using the same procedure, which consists of the following steps:

1. Raw thermal-radiation spectra are acquired immediately after placing the sample on the heater (Fig.S7.1a).
2. The measured spectra are calibrated (Fig.S7.1b) to remove system-dependent contributions and the impact of background.

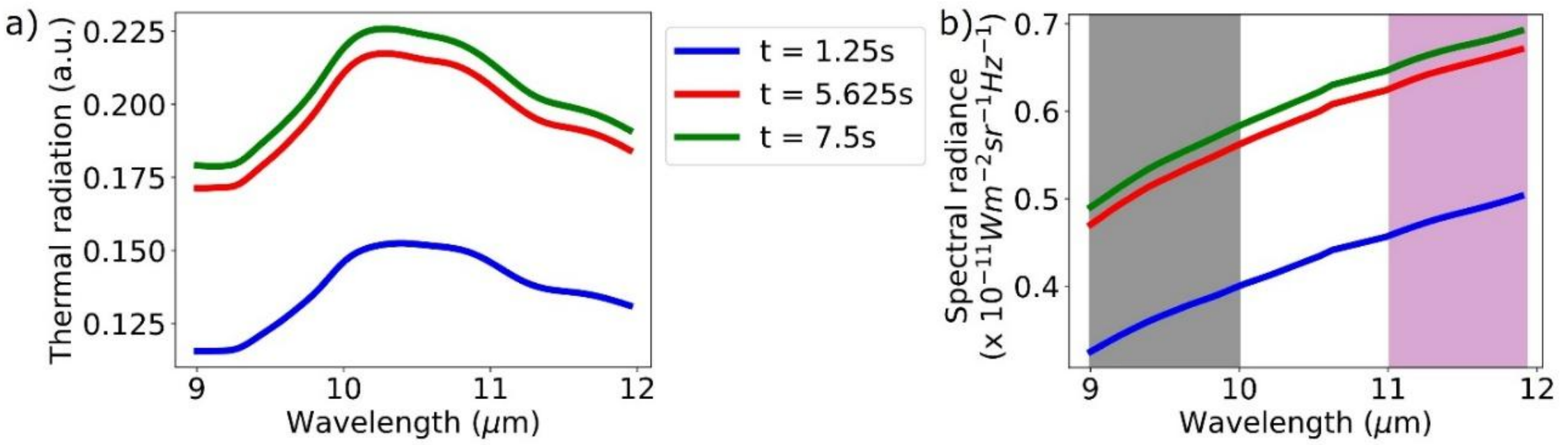

**Fig. S7.1.** Steps 1 and 2: acquisition and calibration of thermal-radiation spectra. Raw (a) and calibrated (b) thermal-radiation spectra measured at three different time points during a single transient heating experiment of a $MgF_2$ window**.** In (b), shaded magenta area indicates the opaque range while shaded grey area indicates the semitransparent range.

3. Transient temperature profiles are inverted using depth thermography principles.
    a. Surface temperature is obtained by fitting the calibrated experimental spectrum in opaque spectral range to the blackbody thermal radiation with unknown temperature multiplied by the emissivity.
    b. Exponential function in the form $T(z) = a + be^{z/c}$ is used for the internal temperature determination. Note that $(a + b)$ = surface temperature.
    c. Unknown coefficients of the exponential function are obtained by fitting the calibrated experimental spectrum in the semitransparent spectral range to the calculated spectrum with a given temperature profile (Fig.S7.2).

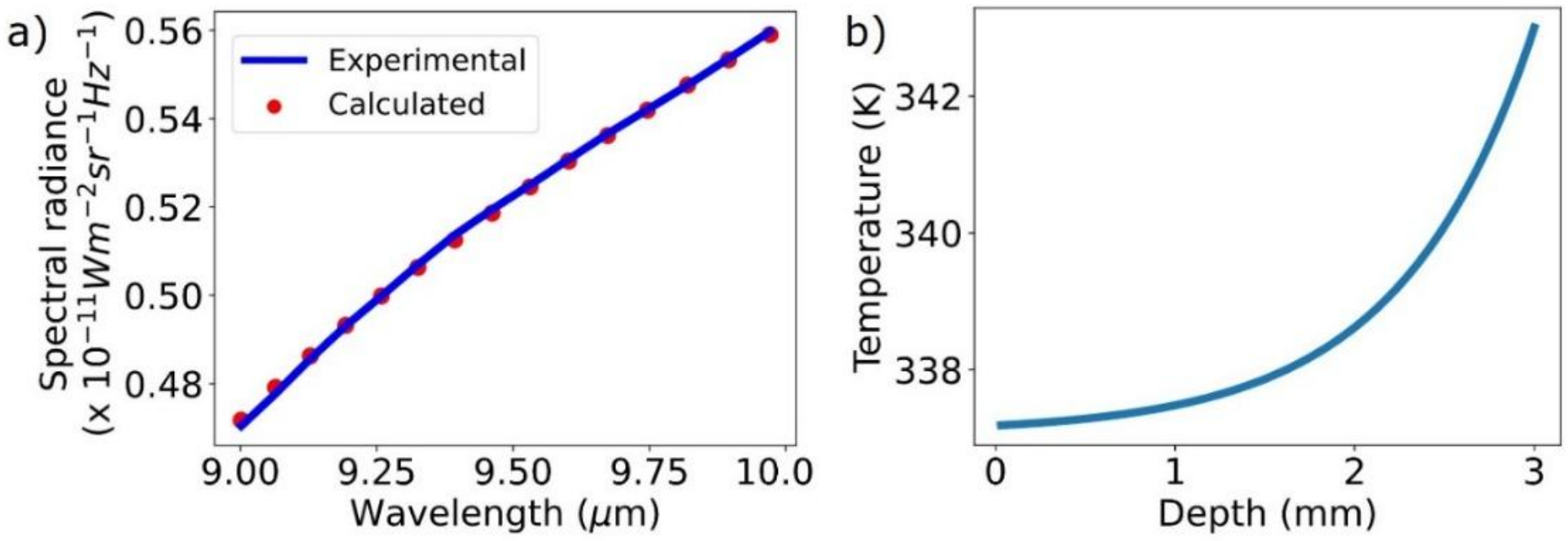

**Fig.S7.2** Temperature inversion (step 3) using transient depth thermography. (a) Measured (solid) and calculated (dotted) using the inverted temperature profile (b) thermal-radiation spectra.

4. Temperature profiles (and corresponding thermal-radiation spectra) close to the steady state are selected (Fig.S7.3).

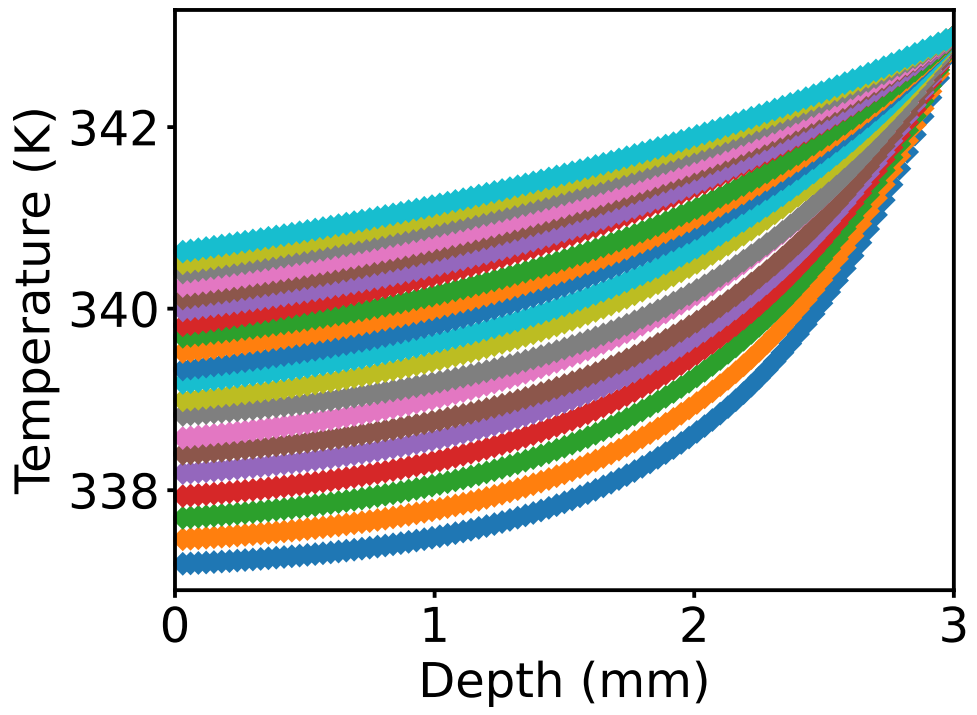

**Fig.S7.3** Step 4**.** First 20 out of 30 selected temperature profiles

5. All possible pairs of temperature profiles where the difference between corresponding thermal-radiation spectra exceeds a threshold value ($0.006*10^{-11}$ $Wm^{-2}sr^{-1}Hz^{-1}$) are considered for thermal conductivity determination.

6. For each considered pair of temperature profiles, a binary search algorithm is implemented to get the “best” thermal conductivity.
    a. A range of 2-8 $Wm^{-1}K^{-1}$ is used with the final resolution of 0.02 $Wm^{-1}K^{-1}$.
    b. A temperature profile/thermal-radiation spectrum corresponding to the earlier time instance is considered as “start”, while the other is “final”.
    c. Using a heat-transfer equation with an initial temperature profile being “start”, the temperature profile (“calculated”) corresponding to the “final” time instance is obtained.
    d. The “calculated” temperature profile is transformed into the thermal-radiation spectrum (“calculated”) using precalculated local emissivities (Fig.S7.4).
    e. The mean squared error between the “calculated” and experimental “final” spectra is being minimized.

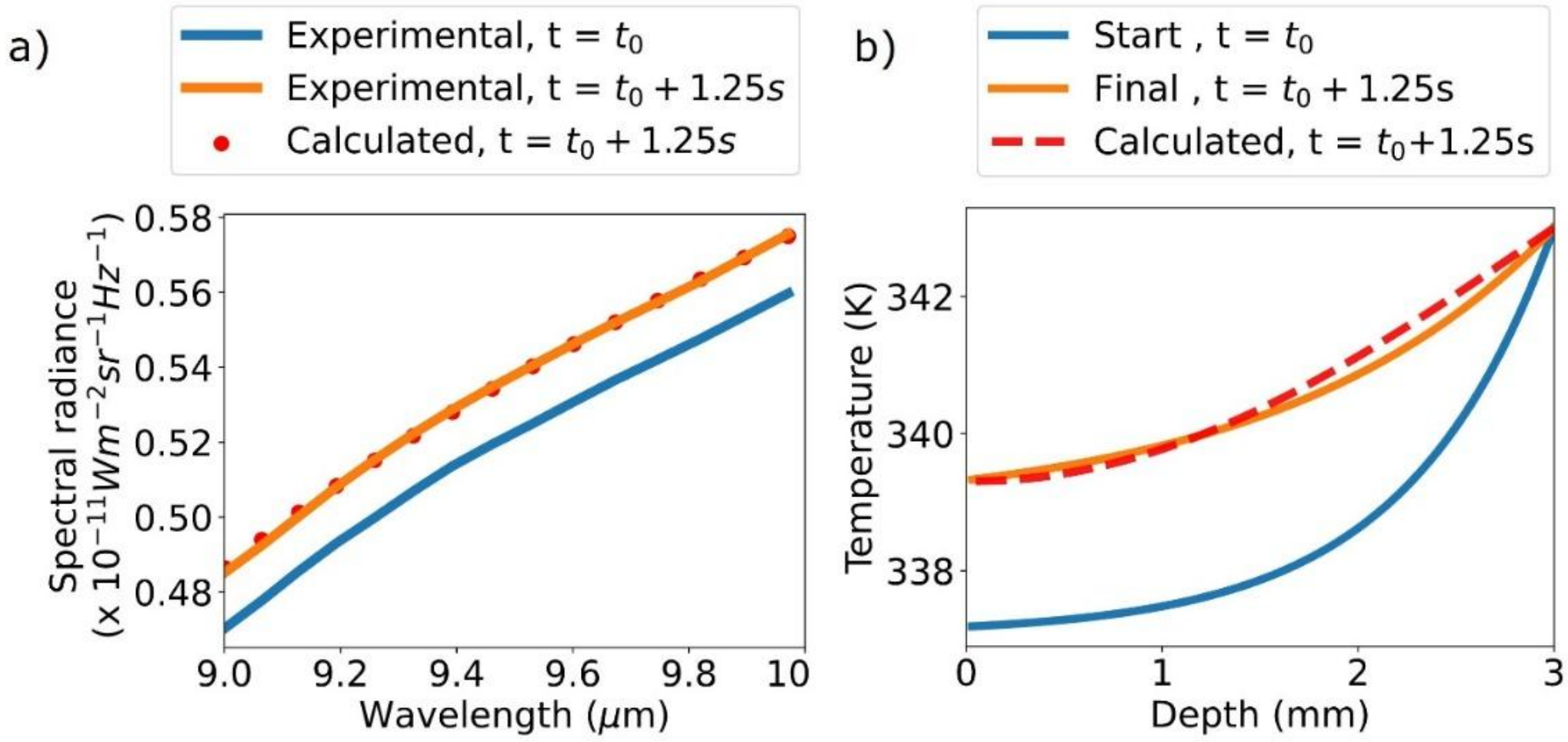

**Fig.S7.4** Step 6. Thermal conductivity determination for a single pair of temperature profiles/thermal-radiation spectra. (a) Measured spectra (solid) – “start” and “final”, and “calculated” using the corresponding temperature profile. (b) Temperature profiles inverted from the thermal-radiation spectra (solid) and the temperature profile (dashed) calculated using the heat transfer equation with k = 5.1 $Wm^{-1}K^{-1}$.

7. The obtained values are averaged over all considered pairs (Fig.S7.5).

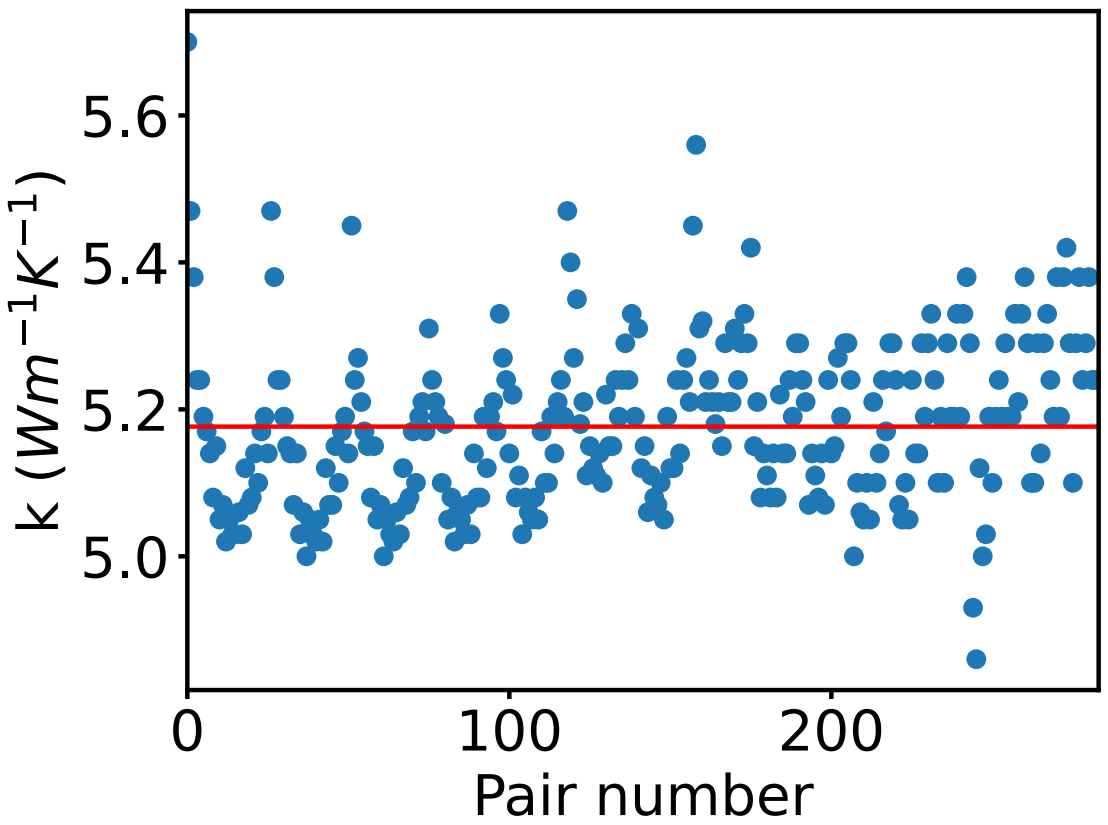

**Fig.S7.5** Step 7**.** Thermal conductivity obtained for different pairs of measurements. Solid red line shows the mean value of 5.17 $Wm^{-1}K^{-1}$.

**Supplement 8. Estimation of natural convection coefficient**.

Convection coefficient ($h$) is related with the Nusselt number ($Nu$) [S5] through the relation $Nu = \frac{hL}{k}$, where $L = \frac{surface\ area}{surface\ perimeter}$ is the characteristic length defined as the ratio of the heated surface area to its perimeter, and $k$ is a thermal conductivity of fluid. For the upper surface of a heated plate, which corresponds to the geometry of our experiment, the following correlations are commonly used [S5]:

$$\overline{Nu_L} = 0.54 Ra_L^{1/4} \quad (if\ Ra_L < 10^7) \text{ and } \overline{Nu_L} = 0.15 Ra_L^{1/3} \quad (if\ Ra_L > 10^7), \tag{S8.1}$$

where $Ra_L$ is a Rayleigh number defined as

$$Ra_L = \frac{g\beta(T_s - T_{room})L^3}{\vartheta\alpha}, \tag{S8.2}$$

where g = 9.8 m/s$^2$ is a gravitational acceleration, $\beta = \frac{1}{T_f} = \frac{2}{(T_s + T_{room})}$ is an inversed effective temperature, $T_s$ is a surface temperature, $T_{room}$ is an ambient room temperature, $\vartheta$ and $\alpha$ are the kinematic viscosity and thermal diffusivity of fluid respectively.

Using these equations as well as parameters of fluid (Table S8.1) from the literature [S5] we estimated that for the samples used in the experiment convection coefficient is $h = 12\ W/(m^2\ K)$ at 100 °C and $h = 14\ W/(m^2\ K)$ at 200 °C.

**Table S8.1 Parameters of air at temperatures corresponding to the plate's temperature of 100°C and 200°C.**

| $T_f$ *(K)* | $\alpha \cdot 10^6$ *(m²/s)* | $\vartheta \cdot 10^6$ *(m²/s)* | $k \cdot 10^3$ *(W/mK)* |
|---|---|---|---|
| 334 | 27.53 | 19.31 | 28.82 |
| 384 | 35.6 | 24.65 | 32.58 |

To determine the thermal conductivity of fused silica, convective heat transfer was taken into account. The convection coefficient and thermal conductivity were determined using the procedure described in Supplementary Note 7. The only modification was that the binary search algorithm was applied over a range of convection coefficient values. For each convection coefficient, the corresponding thermal conductivity was determined, and the pair that minimized the difference between the final experimental and calculated spectra was selected as the optimal solution for each temperature profile/thermal-radiation spectrum pair. The resulting convection coefficient and thermal conductivity values were then averaged over all datasets. The determined convection coefficients at the investigated temperatures are plotted in Fig. S8.1 and show good agreement with the theoretically predicted values.

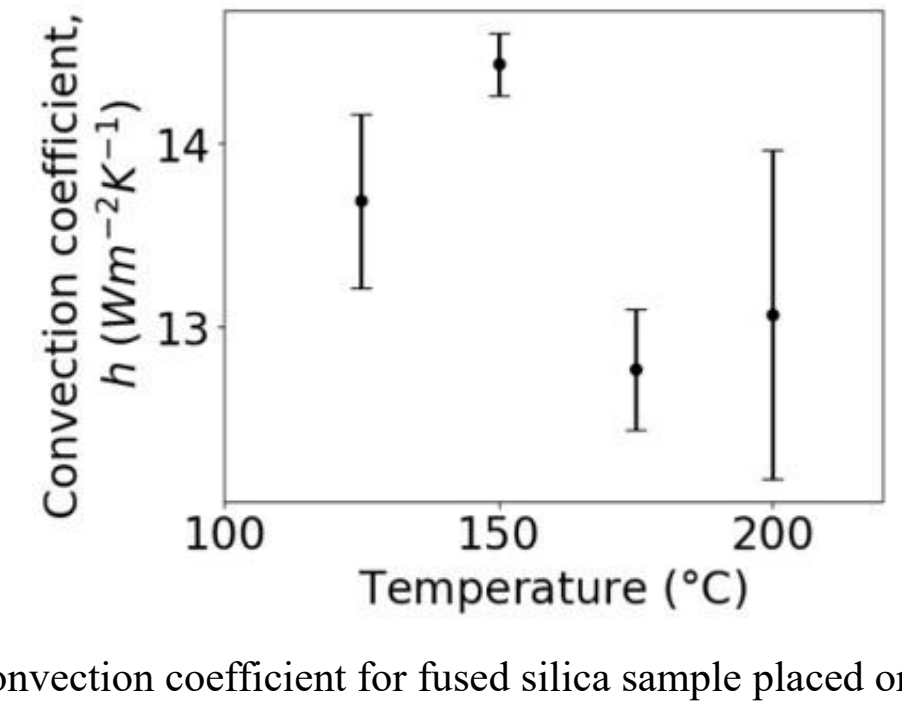

**Fig.S8.1** Experimentally obtained values of convection coefficient for fused silica sample placed onto the heater at different temperatures.

**Supplement 9. Hot-disk measurements.**

We measured the room-temperature thermal conductivity of fused silica using a Thermal Constant Analyzer system from ThermTest Inc., which involved a TPS 1500 module, a dynamic Kapton sensor and a computer equipped with the Hot Disk Thermal Constants analyzer software (version 7.1.16). The machine detects the thermal conductivity in the range of 0.01–400 W/m·K. The dynamic Kapton sensor was sandwiched between two fused-silica samples. The device measures thermal conductivity using the transient plane source method, in which the sensor functions as both a heat source and a resistance thermometer. An electrical heating pulse is applied, and the resulting temperature rise is recorded over time. Due to the sample size requirements for the sensor, 5-mm-thick fused-silica windows purchased from Thorlabs were used for these measurements. The values of thermal conductivities measured with various heating powers and measurement durations are shown in Fig. S9.1. The mean value of the measured thermal conductivity as well as standard deviation are shown in the main manuscript in the Fig. 3.f.

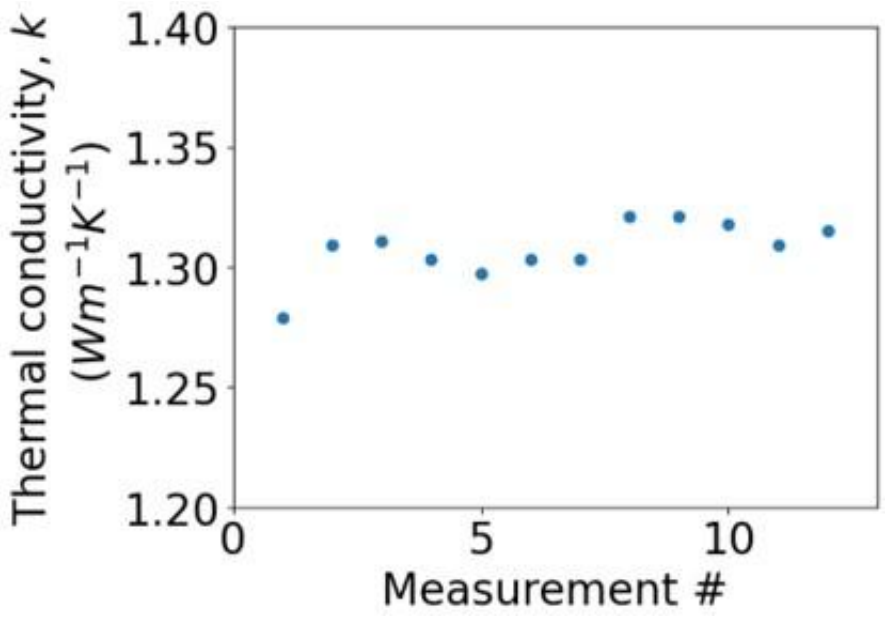

**Fig.S9.1** Thermal conductivity of fused silica obtained using the Hot Disk TPS 1500 system at room temperature. Different measurements correspond to various combinations of heating power and measurement time.

**Supplement 10. Temperature-dependent thermal conductivity.**

In the thermal conductivity determination procedure (Supplementary Note 7), we used temperature profiles that were close to steady state. Applying the same procedure to earlier transient measurements, however, reveals several notable observations. First, the extracted thermal conductivity depends strongly on the selected pair of temperature and thermal-radiation profiles. Second, this variation follows the intrinsic temperature dependence of the material. For fused silica, the extracted thermal conductivity increases with time, whereas for $MgF_2$ it decreases (Fig. S10.1 where the heater temperature was set to 200 °C). Note that the red rectangular in (a) and (c) shows the pairs of combinations used for extracting thermal conductivity around 200°C.

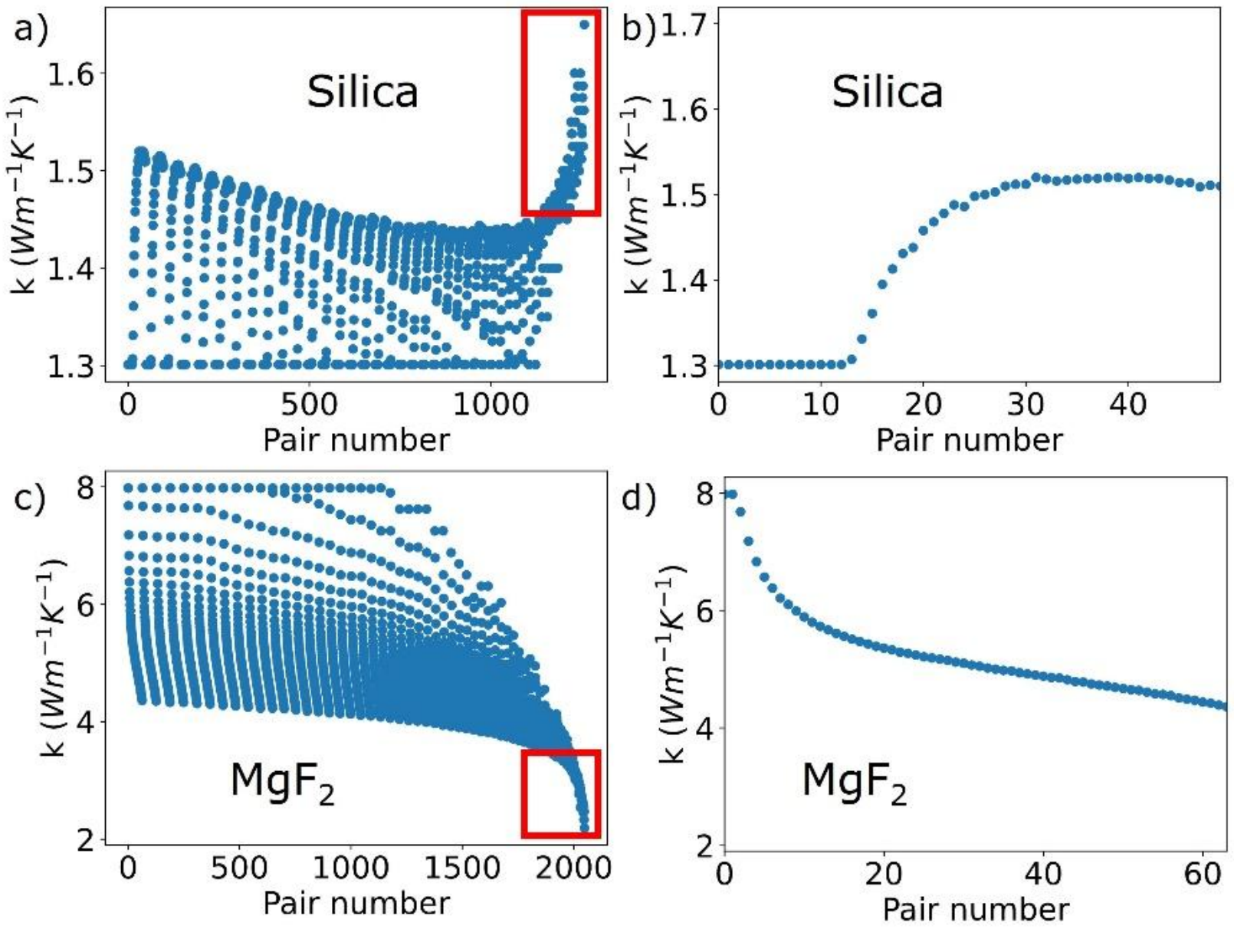

**Fig.S10.1** Thermal conductivity search includes pairs of temperature profiles/thermal-radiation spectra far from the steady-state for (a, b) fused silica and (c, d) $MgF_2$ samples. The red rectangles indicate the pairs used for the actual thermal conductivity determination. Pair numbers are assigned sequentially along the measurement time: the first pair is formed by measurements 1 and 2, the second pair is formed by measurements 1 and 3, while the last pair consists of the penultimate and final measurements. (b) and (d) include pairs formed by the first and later measurements. Ascending (b) and descending (d) trends can be observed for fused silica and $MgF_2$ respectively.

Figures S10.1b and S10.1d clearly illustrate the increasing and decreasing trends, respectively. In these plots, each measurement pair consists of one reference (permanent) measurement and a subsequent measurement acquired at a later time. For $MgF_2$ (Fig. S10.1d), the thermal conductivity decreases with increasing pair number, indicating that the thermal conductivity decreases with temperature, since a larger pair number corresponds to a higher average temperature. In contrast, fused silica exhibits the opposite trend, with thermal conductivity increasing as the average temperature rises. These observations suggest that transient thermal-radiation spectra could be used to directly determine the temperature dependence of thermal conductivity from a single transient measurement, potentially eliminating the need for multiple experiments at different temperatures.

**Supplement 11. Heat transfer simulations with thermal boundary resistance.**

In the simulations of heat transfer within the two-layer structure with thermal boundary resistance, we considered GaN and AlGaN as the constituent materials. The material properties of AlGaN depend strongly on the Al fraction, while the properties of GaN can vary considerably from sample to sample. Therefore, we used density, heat capacity, and thermal conductivity values within the ranges reported in the literature [S6–S8]. The material properties used for GaN and AlGaN are summarized in Table S11.1.

**Table S11.1 Material properties of GaN and AlGaN.**

| Material | Density (kg/m3) | Heat capacity (J/kg K) | Thermal conductivity (W/m K) |
|---|---|---|---|
| GaN | 6100 | 490 | 150 |
| AlGaN | 4200 | 700 | 10 |

In the simulations, we assumed a constant heat flux incident on the GaN surface. The temperature of the top (first) layer of the discretized GaN evolves with time according to

$$T_1^{n+1} = T_1^n + \frac{\Delta t}{\rho_{GaN} c_{p_{GaN}} \Delta x}\left(q - k_{GaN}\frac{T_1^n - T_2^n}{\Delta x}\right), \quad \text{(S11.1)}$$

where $T_i^j$ is a temperature of layer $i$ at time step $j$, $\Delta t$ is the time step, $\Delta x$ is the spatial step, $q$ is the incident heat flux, and $\rho, c_p,$ and $k$ are the density, heat capacity and thermal conductivity of the material, respectively. For a homogeneous material, temperature $T_i^j$ for a layer $i$ within the sample evolves according to

$$T_i^{n+1} = T_i^n + \frac{\alpha \Delta t}{\Delta x^2}(T_{i+1}^n - 2T_i^n + T_{i-1}^n), \quad \text{(S11.2)}$$

where $\alpha = \frac{k}{\rho c_p}$ is thermal diffusivity. The treatment of the interface between two different materials is more complicated. Here, we account for the interface using the concept of effective thermal boundary resistance, which can be expressed as [S9]

$$G^{-1} = \frac{\Delta x}{2k_{GaN}} + R + \frac{\Delta x}{2k_{AlGaN}}, \quad \text{(S11.3)}$$

where $R$ is a thermal boundary resistance (TBR). Accordingly, the temperature of the bottom GaN layer ($T_m^n$) is given by

$$T_m^{n+1} = T_m^n + \frac{\alpha_{GaN}\Delta t}{\Delta x^2}(T_{m-1}^n - T_m^n) + \frac{G\Delta t}{\rho_{GaN} c_{p_{GaN}} \Delta x}(T_{m+1}^n - T_m^n), \quad \text{(S11.4)}$$

while the temperature of the top AlGaN layer ($T_{m+1}^n$) is given by

$$T_{m+1}^{n+1} = T_{m+1}^n + \frac{\alpha_{AlGaN}\Delta t}{\Delta x^2}(T_{m+2}^n - T_{m+1}^n) + \frac{G\Delta t}{\rho_{AlGaN} c_{p_{AlGaN}} \Delta x}(T_m^n - T_{m+1}^n) \quad \text{(S11.5)}$$

In the limiting case of a homogeneous material with no thermal boundary resistance ($R = 0$), Eqs. S11.4 and S11.5 reduce to Eq. S11.2.

Using the equations above, we calculated transient temperature profiles within the GaN/AlGaN structure for several values of the thermal boundary resistance (Fig. S11.1). A value of $R = 2 \times 10^{-9}\ m^2K/W$ is a typical value of TBR for a well-manufactured GaN/AlGaN interface [S10]. The two higher values considered here represent poorer-quality interfaces, as evidenced by the larger temperature gradient across the interface (Fig. S11.1a, b).

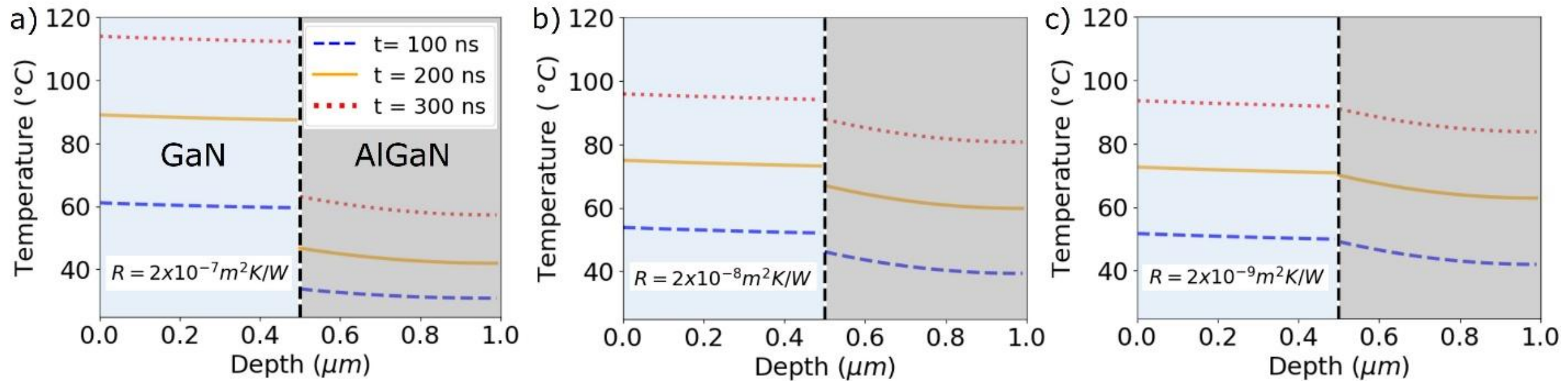

**Fig.S11.1** Calculated temperature profiles inside GaN/AlGaN sample heated by a CW laser with a flux of 700 $MW/m^2$ at three time instances. The interface between GaN and AlGaN has a thermal boundary resistance of (a) $R = 2 \times 10^{-7}\ m^2K/W$, (b) $R = 2 \times 10^{-8}\ m^2K/W$ and (c) $R = 2 \times 10^{-9}\ m^2K/W$.

To calculate the thermal radiation spectrum emitted by a heated GaN/AlGaN composite, it is necessary to determine the local emissivity, which in turn depends on the optical properties of the constituent materials. The real and imaginary parts of the refractive indices of GaN and AlGaN were taken from [S11] and [S12], respectively, and are shown in Fig. S11.2.

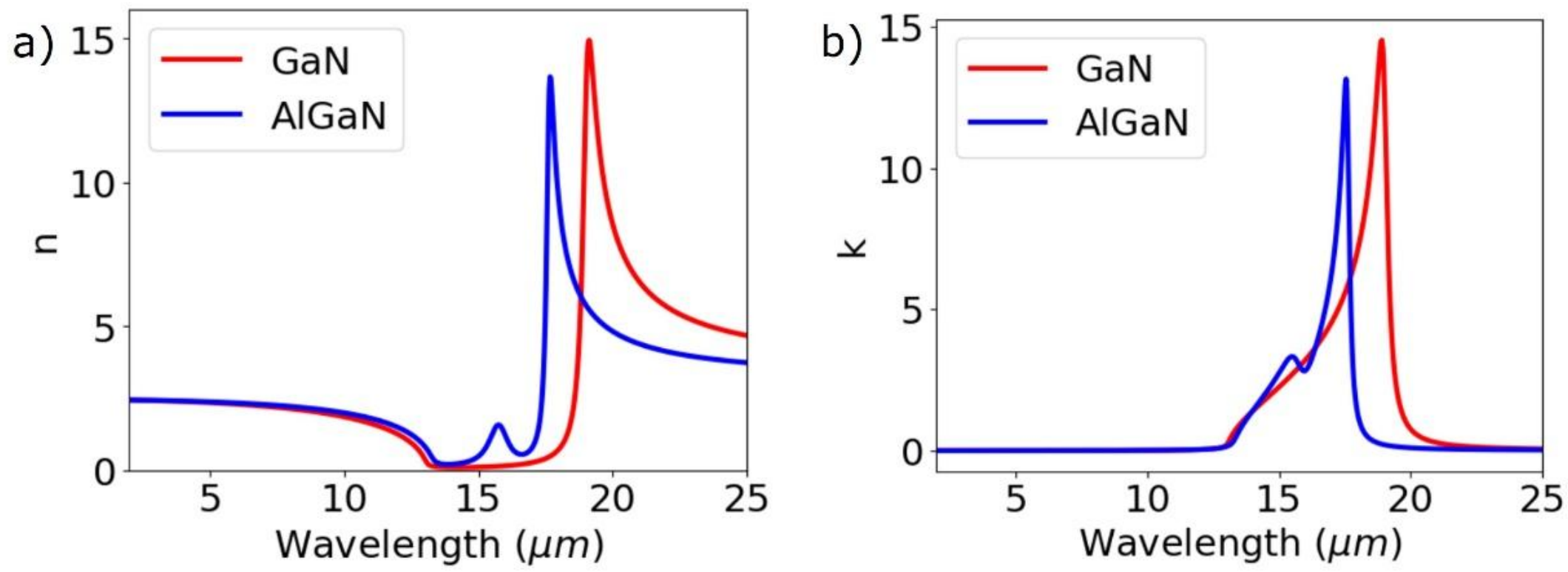

**Fig.S11.2** Optical properties of GaN and AlGaN, (a) Real and (b) imaginary parts of the refractive index.